\documentclass[preprint,12pt,authoryear]{elsarticle}

\usepackage{lineno,hyperref}
\usepackage[utf8]{inputenc}
\usepackage[T1]{fontenc}
\usepackage[abs]{overpic} 
\usepackage{amssymb}
\usepackage{xcolor}
\usepackage{upgreek}
\usepackage[normalem]{ulem}

\modulolinenumbers[5]

\journal{Advances in Space Research}

\biboptions{authoryear}

\begin{document}

\begin{frontmatter}

\title{Radiation Monitor RADMON aboard Aalto-1 CubeSat: First results}


\author[UTUF]{Jan Gieseler\corref{mycorrespondingauthor}}
\cortext[mycorrespondingauthor]{Corresponding author}
\ead{jan.gieseler@utu.fi}
\author[UTUF]{Philipp Oleynik}
\ead{philipp.oleynik@utu.fi}
\author[UTUF]{Heli Hietala}
\ead{heli.hietala@utu.fi}
\author[UTUF]{Rami Vainio}
\ead{rami.vainio@utu.fi}
\author[UTUT]{Hannu-Pekka Hedman}
\ead{hannu-pekka.hedman@utu.fi}
\author[UTUF]{Juhani Peltonen}
\ead{juhpe@utu.fi}
\author[UTUF]{Arttu Punkkinen}
\ead{arjupu@utu.fi}
\author[UTUT]{Risto Punkkinen}
\ead{rpunk@utu.fi}
\author[UTUT]{Tero S\"antti}
\ead{teansa@utu.fi}
\author[Hel]{Edward H{\ae}ggstr\"om}
\ead{edward.haeggstrom@helsinki.fi}
\author[Aalto]{Jaan Praks}
\ead{jaan.praks@aalto.fi}
\author[Aalto]{Petri Niemel\"a}
\ead{petri.niemela@aalto.fi}
\author[Aalto]{Bagus Riwanto}
\ead{bagus.riwanto@aalto.fi}
\author[Aalto]{Nemanja Jovanovic}
\ead{nemanja.jovanovic@aalto.fi}
\author[Aalto]{M.~Rizwan Mughal}
\ead{muhammad.2.mughal@aalto.fi}

\address[UTUF]{Department of Physics and Astronomy, University of Turku, \\
20014 University of Turku, Finland}
\address[UTUT]{Department of Future Technologies, University of Turku,\\ 
20014 University of Turku, Finland}
\address[Hel]{Department of Physics, University of Helsinki, \\
PO Box 64, 00014 University of Helsinki, Finland}
\address[Aalto]{School of Electrical Engineering, Aalto University,\\ 
PO Box 11000, 00076 Aalto, Finland}

\begin{abstract}
The Radiation Monitor (RADMON) on-board Aalto-1 CubeSat is an energetic particle detector that fulfills the requirements of small size, low power consumption and low budget. Aalto-1 was launched on 23 June 2017 to a sun-synchronous polar orbit with 97.4$^\circ$ inclination and an average altitude of somewhat above 500 km. RADMON has been measuring integral particle intensities from October 2017 to May 2018 with electron energies starting at low-MeV and protons from 10~MeV upwards. In this paper, we present first electron and proton intensity maps obtained over the mission period. In addition, the response of RADMON measurements to magnetospheric dynamics are analyzed, and the electron observations are compared with corresponding measurements by the PROBA-V/EPT mission. Finally, we describe the RADMON data set, which is made publicly available.
\end{abstract}

\begin{keyword}
radiation belts\sep 
magnetosphere\sep
electron precipitation\sep
energetic particle telescope\sep
CubeSat
\end{keyword}

\end{frontmatter}


\section{Introduction}
The dynamics of the radiation environment of the Earth \citep[e.g.,][]{Vainio-etal-2009} is one of the scientific targets that can be successfully addressed by small satellites, such as CubeSats \citep[e.g.,][]{Li-etal-2013,Li-etal-2017}. In this paper we will report on energetic charged particle radiation measurements with a simple instrument on-board a CubeSat in Low Earth Orbit (LEO), scanning the radiation belts and making successful observations on their dynamics. We present the observations and compare them with another instrument at somewhat higher orbit demonstrating good correspondence between the two at electron energies above 1 MeV. The purpose of the paper is to present the measurement data and to release the data set to the community for further exploitation.

\section{Aalto-1 mission}
\subsection{Aalto-1 CubeSat}
\label{sect:Aalto1cubesat}
Aalto-1 \citep{Kestila-etal-2013} is a three-unit, 4-kg CubeSat launched on 23 June 2017 to a sun-synchronous polar orbit with 97.4$^\circ$ inclination and an average altitude somewhat above 500 km \citep{Praks-etal-2018}. It is Finland's first national satellite mission, and has scientific, technology demonstration, and educational objectives. The satellite carries three payloads, a hyper-spectral camera \citep{Kestila-etal-2013, Praks-etal-2018}, a Plasma Brake -- a de-orbiting device made of a thin charged tether interacting with the ionospheric plasma when deployed \citep{Janhunen-2010} -- and a Radiation Monitor (RADMON) sensitive to $>$10~MeV protons and $>$1~MeV electrons. Apart from the payloads, the satellite consists of several subsystems including the electrical power system, the attitude determination and control system (ADCS), a two-channel radio link and a Linux-based on-board computer. The satellite platform and RADMON have been produced by teams of students at Aalto University, University of Turku, and University of Helsinki.

After commissioning of the subsystems, the Aalto-1 mission was divided into slots for each of the payloads. RADMON observations were concentrated on the fall-winter period of 2017, with some additional slots during spring 2018. In this paper we will concentrate in particular on the period from early October to late December, when the instrument was observing the radiation belts.

After launch, the spacecraft was in a tumbling state, while several attempts to stabilize the spacecraft with magnetorquers were made. The ADCS magnetometer and gyros have been used to determine the rotation rates around the three spacecraft axes. Since RADMON has its view cone directed along the long $Z$ axis of the spacecraft, the projection of the angular velocity to the $XY$ plane, $\Omega_{XY}$, mainly determines the rate at which RADMON samples the directions in the sky. 
The rotation rate and rotation axis of the spacecraft changes slowly over the mission period due to the disturbance torques from the environment, but until mid October 2017 typical values of a few degrees per second were observed for $\Omega_{XY}$, while at the same time the rotation rate around $Z$ axis was large, more than 50$^\circ$/s, on average. This means that the spacecraft is rotating mainly around the $Z$ axis. In the inertial frame, the spacecraft is mainly rotating around its $Z$ axis, as well, but in addition this axis precesses around the direction of angular momentum at rates of about (10$\pm$4)$^\circ$/s, as determined from the magnetometer data of the ADCS. Some nutation seems to be present as well. The precession rate mainly determines the rate at which RADMON scans the Sky.
Unfortunately, the ADCS data is under-sampled so the actual attitude of the spacecraft at each time cannot be derived, even if the rotation rates can be obtained. From mid October until mid November, there is even no rates data from the ADCS system. 
The rotation rate $\Omega_{XY}$ nevertheless increases over this period, and the increased rate of $\Omega_{XY}$ is accompanied with an increase of the precession rate to $\sim$30$^\circ$/s in mid November. Starting from 12 November until 15 December 2017, $\Omega_{XY}$ then increases from about 17$^\circ$/s to about 37$^\circ$/s and from mid November until the end of the main RADMON measurement period the rate at which the instrument scans the sky is large enough to fit more than one rotation in the integration time of 15 seconds.
Thereafter, $\Omega_{XY}$ increases more slowly to about 60$^\circ$/s in early February 2018. The changes in $\Omega_{XY}$ are mainly related to the change of the main axis of rotation from the $Z$ axis to the $Y$ axis.

\subsection{RADMON instrument}
RADMON \citep{Peltonen-etal-2014} is a small (0.4U, 360~g), low-power (1~W) and low-cost energetic particle instrument consisting of four subsystems: the detector unit, the analog board, the digital board, and the power-supply unit, integrated in a compact stack electrically connected via the instrument bus (a 52-pin PC/104 connector) and directly integrated to the spacecraft Printed Circuit Board (PCB) stack. The detector unit contains two detectors, a small (2.1$\times$2.1$\times$0.35~mm$^3$) Si detector and a larger cubic (1~cm$^3$) CsI(Tl) scintillator with a photodiode readout. The detectors form a stack producing two energy-loss signals per a detected particle, which are used in the standard $\Delta E$--$E$ analysis to derive the deposited total energy and the particle species. 
A 280 $\upmu$m aluminum window on top of the detector stack together with the detection logic of the instrument sets the energy threshold for electrons to about 1~MeV and protons to about 10~MeV. This matter is discussed in detail by \citet{Oleynik-etal-2019}.

The analog board contains the bias generators, the analog amplifiers and the analog-to-digital converters (ADC) for the two channels that are continuously sampled at 10~MHz rate. The digital board performs all the signal processing from pulse detection to energy-loss determination and particle classification into the different spectral channels. 
The basic time resolution of the instrument was chosen to be 15 seconds, yielding a good compromise between counting statistics and spatial resolution corresponding to about 1~degree in latitude on the polar orbit of the spacecraft.
The particle counting rates are delivered in nine nominal proton and five nominal electron channels. The responses of these channels are reported in detail by \citet{Oleynik-etal-2019}, where also a more detailed description of the instrument is presented. 

While the electron energy channels of RADMON are quite integral in nature, especially the channels e2 and e3 respond primarily to electrons with cutoff energies of about 1.5 and 3.1~MeV, respectively. Channel e4 has a higher cutoff energy (6.2~MeV) and e5 responds primarily to very high-energy protons. Channel e1 is very narrow, just below e2, and has very few counts. This channel has practically been eliminated because of the need to increase the detection threshold of the scintillator channel that picks up noise from somewhere in the spacecraft system.

The first four proton channels are aimed at providing a measurement of differential proton spectrum at 10--40~MeV energies with relatively good energy resolution in the nominal range, but with some high-energy side bands that render their response rather integral. The remaining channels p5 to p9 have more involved responses but can be combined into a single channel at about 40--80~MeV with well defined response.

\section{RADMON data}
\label{sect:RADMON-data}
\begin{figure}
\includegraphics[trim=0 40 0 0, clip, width=\columnwidth]{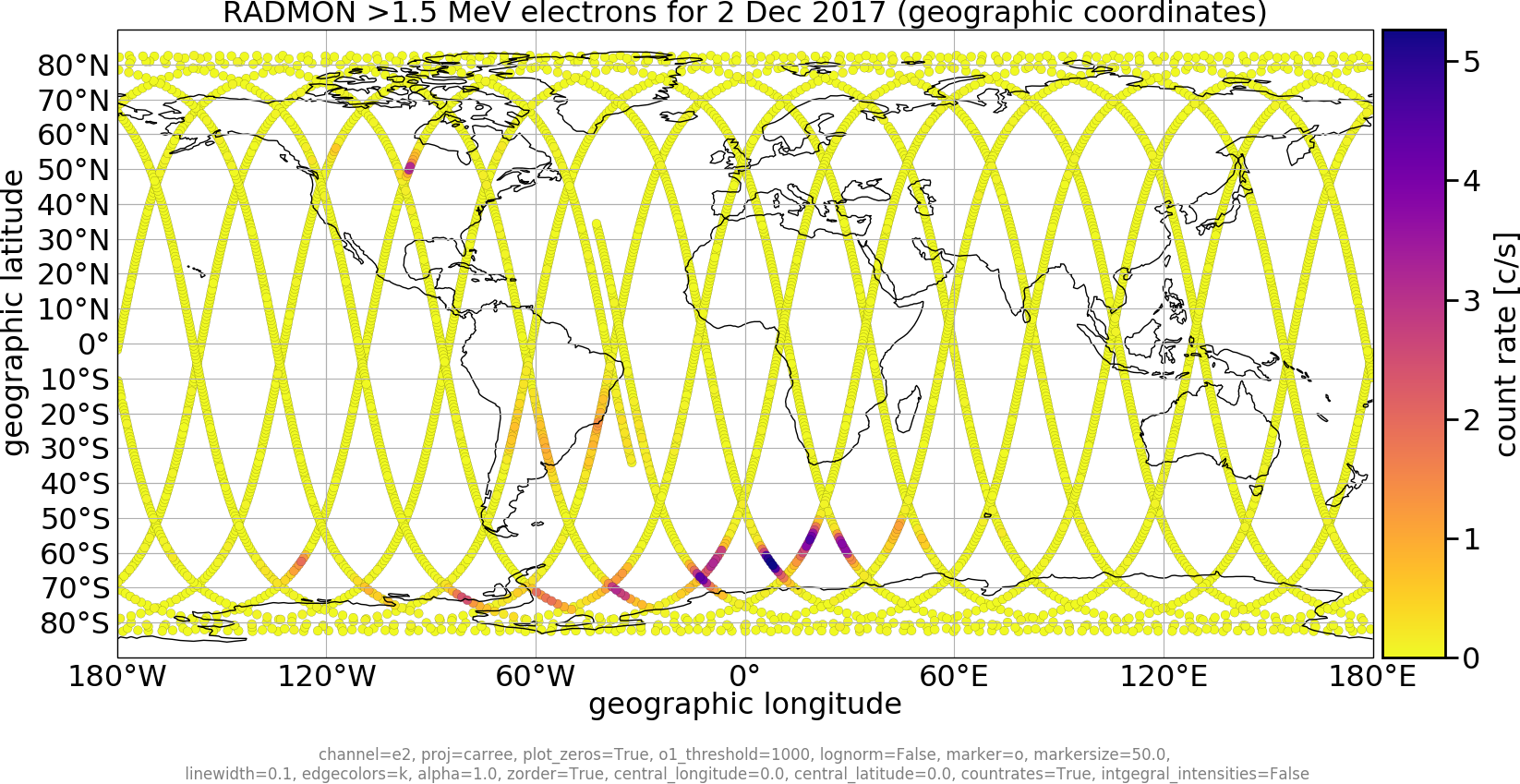}
\caption{Example of RADMON electron count rate measurements in channel e2 ($>$1.5~MeV) along the Aalto-1 orbit during one day (2 Dec 2017).}
\label{fig:2017-12-02_plasma_crop}
\end{figure}
\begin{figure}
\includegraphics[width=\columnwidth]{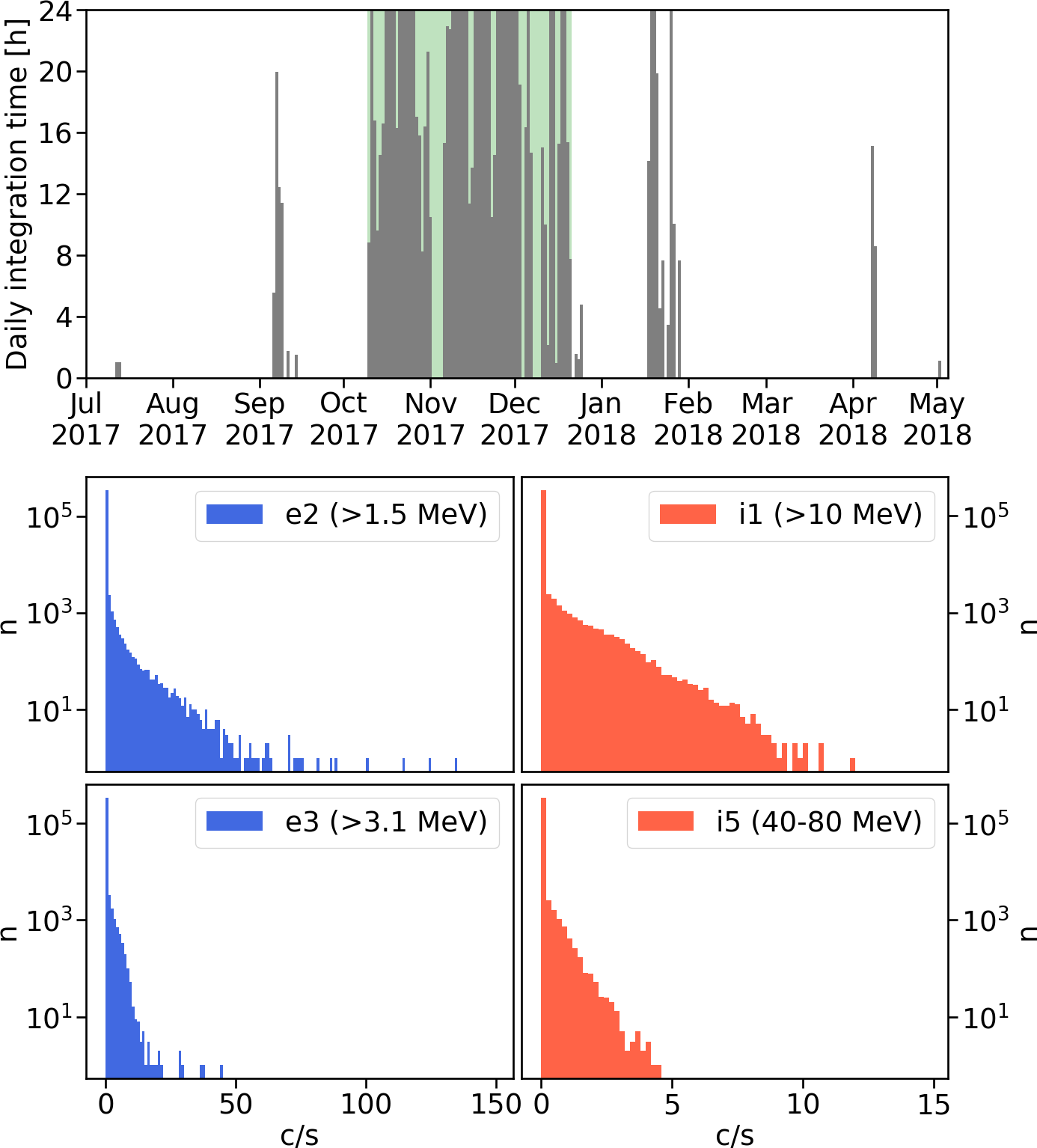}
\caption{Top: Daily integrated measurement time of RADMON in hours (x-ticks mark the first day of each month) with green shading indicating the main measurement period analyzed later. In general, 10 Oct 2017 is regarded as the start point of the mission period because at previous times RADMON calibration was not final. Bottom: Count rate histograms for selected electron (left) and combined proton (right) channels (i$1=\sum_{i=1}^{9} p_i$ and i$5=\sum_{i=5}^{9} p_i$) for the time range Oct 2017 to May 2018.}
\label{fig:hist_inttime_channels}
\end{figure}

Because of the rotational state of the satellite, combining multiple RADMON measurements (i.e., the sum of multiple 15-seconds measurement intervals) corresponds to omni-directional (angle-averaged) intensities over given locations. This has been verified by analyzing the ADCS magnetometer measurements. Although these data are only available for some short time intervals, they allow to investigate the temporal variation of the pitch angle of the boresight direction of RADMON. The pitch-angle coverage depends on the relative orientation of the spacecraft $Z$ axis and the local magnetic field, and cannot be derived in a deterministic way. Analysis of the scarce magnetometer data reveals different patterns: Sometimes RADMON covers the whole pitch-angle range during every 15-seconds interval but sometimes it scans a more limited and varying band of pitch-angle cosine values which is changing over a time range of about 15--30 minutes. Therefore in this work RADMON intensities are only used as averages of dozens or more of 15-seconds measurement intervals (see Fig.~\ref{fig:lm_n_series_hist_radmon_e2_crop}). The lack of pitch-angle values for the measurements also means that RADMON measurements cannot be analyzed in terms of the adiabatic invariants. Thus, we will present the observations as a function of coordinates and time in fixed energy channels.

An example of RADMON measurements along the Aalto-1 orbit during one single day (2 December 2017) is shown in Fig.~\ref{fig:2017-12-02_plasma_crop}, depicting the counting rates of $>$1.5~MeV electrons detected in channel e2. Clearly visible are already the higher rates at higher latitudes in both hemispheres. In addition, some increase in the South Atlantic anomaly (SAA) is also indicated. Note that the rate of scanning the sky during this measurement was about 30$^\circ$\,s$^{-1}$, so the counting rates correspond relatively well to angular averages even at the 15 s integration time of the instrument.

\begin{figure}
\includegraphics[trim=0 40 0 0, clip, width=\columnwidth]{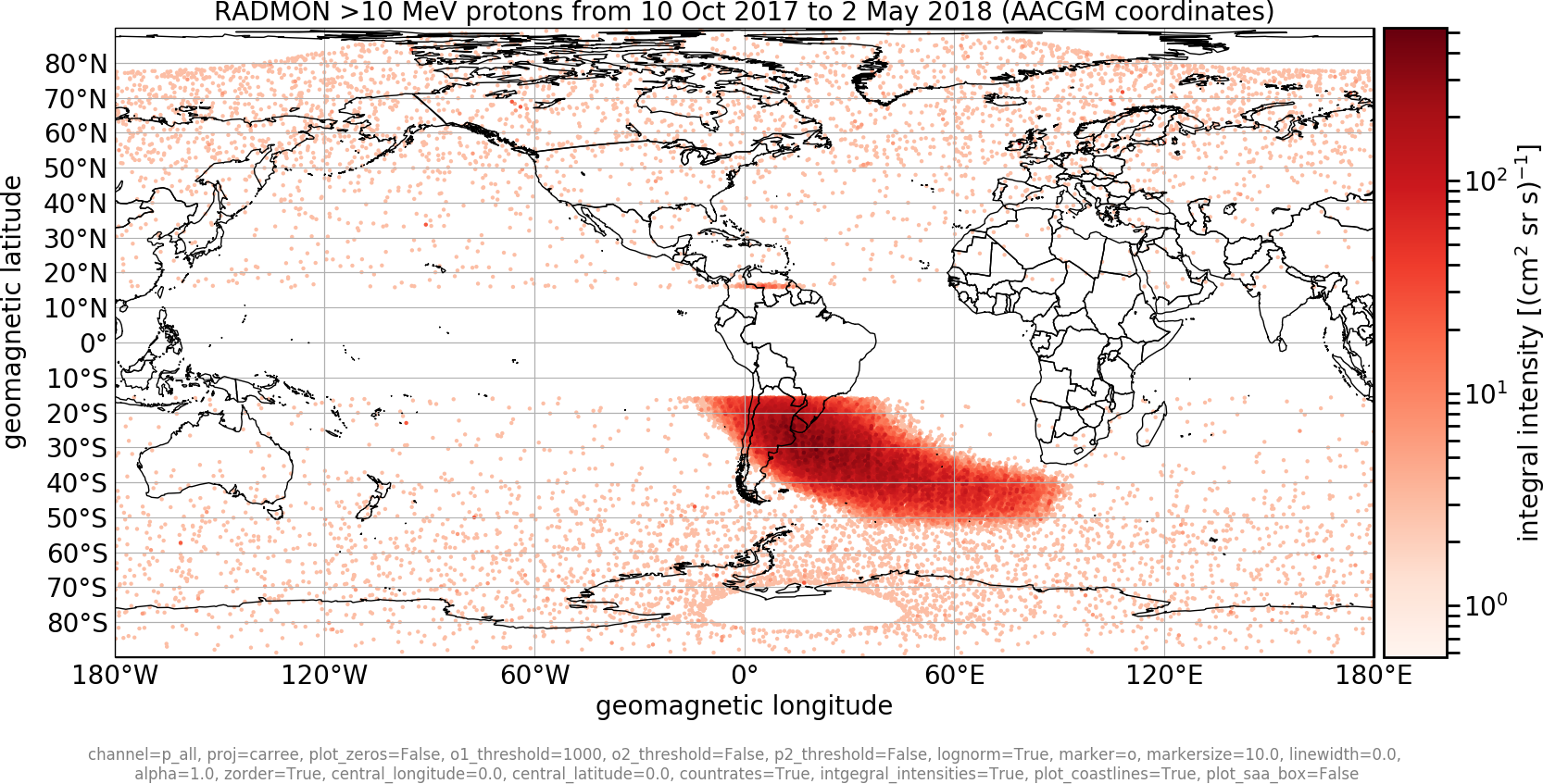}
\caption{Integral intensities of all proton channels combined (corresponding to $>$10~MeV) for the main RADMON mission period from 10 Oct 2017 to 2 May 2018 (omitting zero counts) in altitude adjusted corrected geomagnetic (AACGM) coordinates (note that the highest z values are plotted 'on top' of the lower values). 
}
\label{fig:p-all_aacgmv2_countorder_red_o1_lt_1000_bigger_crop}
\end{figure}

While Aalto-1 satellite was launched already in June 2017, first longer measurements with RADMON took place only in September 2017. During the first observation campaign RADMON calibration was not final. Thus, 10 October 2017 is regarded as the start point of the mission period.
In Fig.~\ref{fig:hist_inttime_channels} (top), the daily integration time of the RADMON instrument is given, with a green shading indicating the main measurement period analyzed later in this work.
Count rate histograms over the whole mission period (October 2017 to May 2018) for two electron and two proton channels are presented in Fig.~\ref{fig:hist_inttime_channels} (bottom). Note that the overall lower proton count rates originate from the fact that protons were only detected in the SAA (cf. also Fig.~\ref{fig:p-all_aacgmv2_countorder_red_o1_lt_1000_bigger_crop}). 
Only in this region the trapped protons of inner Van Allen radiation belt reach to lower altitudes (due to the weak magnetic field there), to the orbit of Aalto-1.

\begin{figure}
\includegraphics[trim=0 40 0 0, clip, width=\columnwidth]{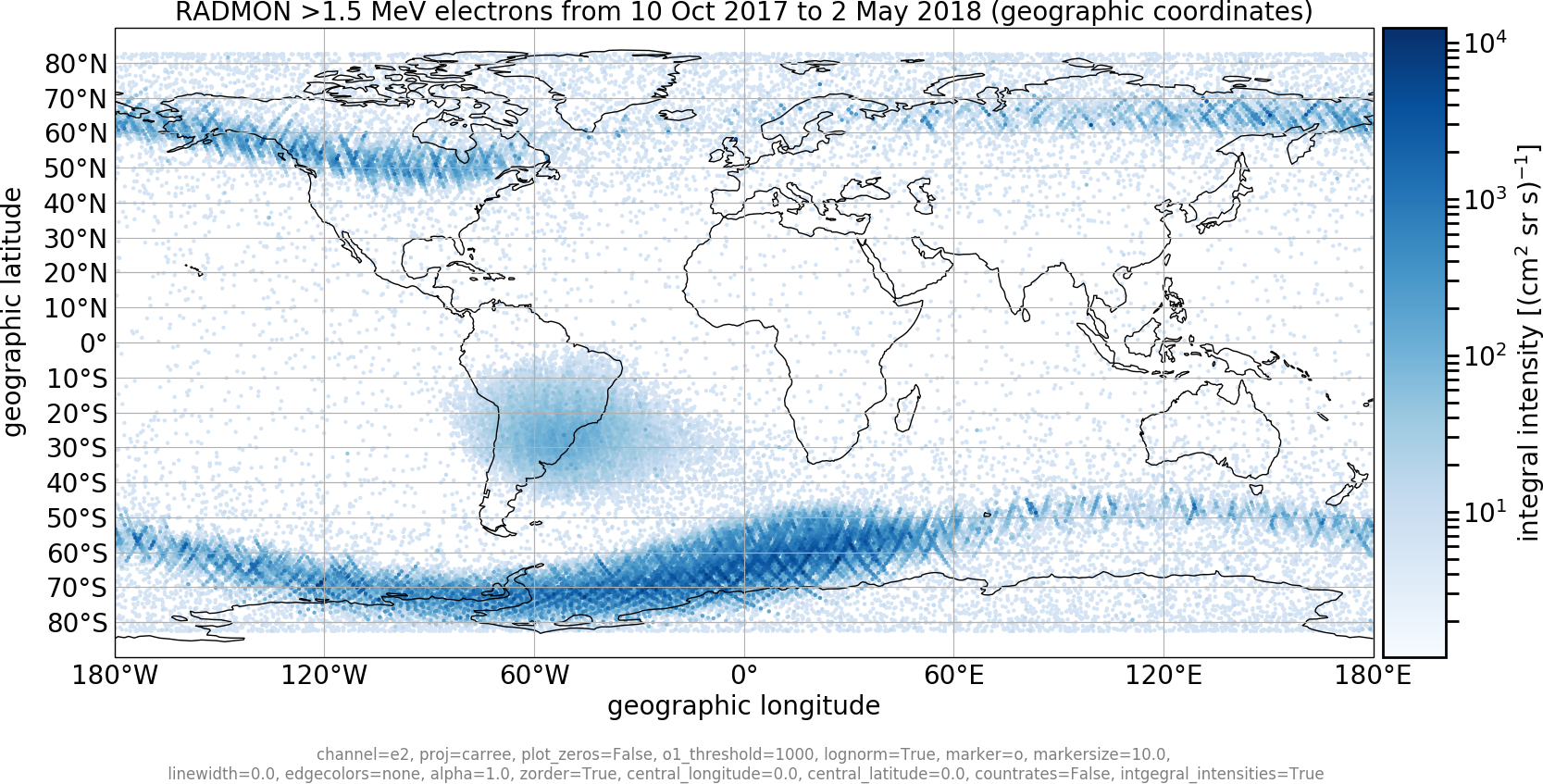}
\caption{Integral intensities of channel e2 (electrons with $>$1.5~MeV) for the main RADMON mission period from 10 Oct 2017 to 2 May 2018 (omitting zero counts) in geographic coordinates (note that the highest z values are plotted 'on top' of the lower values).}
\label{fig:e2_countorder_blue_o1_lt_1000_bigger_crop}
\end{figure}

\begin{figure}
\includegraphics[trim=0 40 0 0, clip, width=\columnwidth]{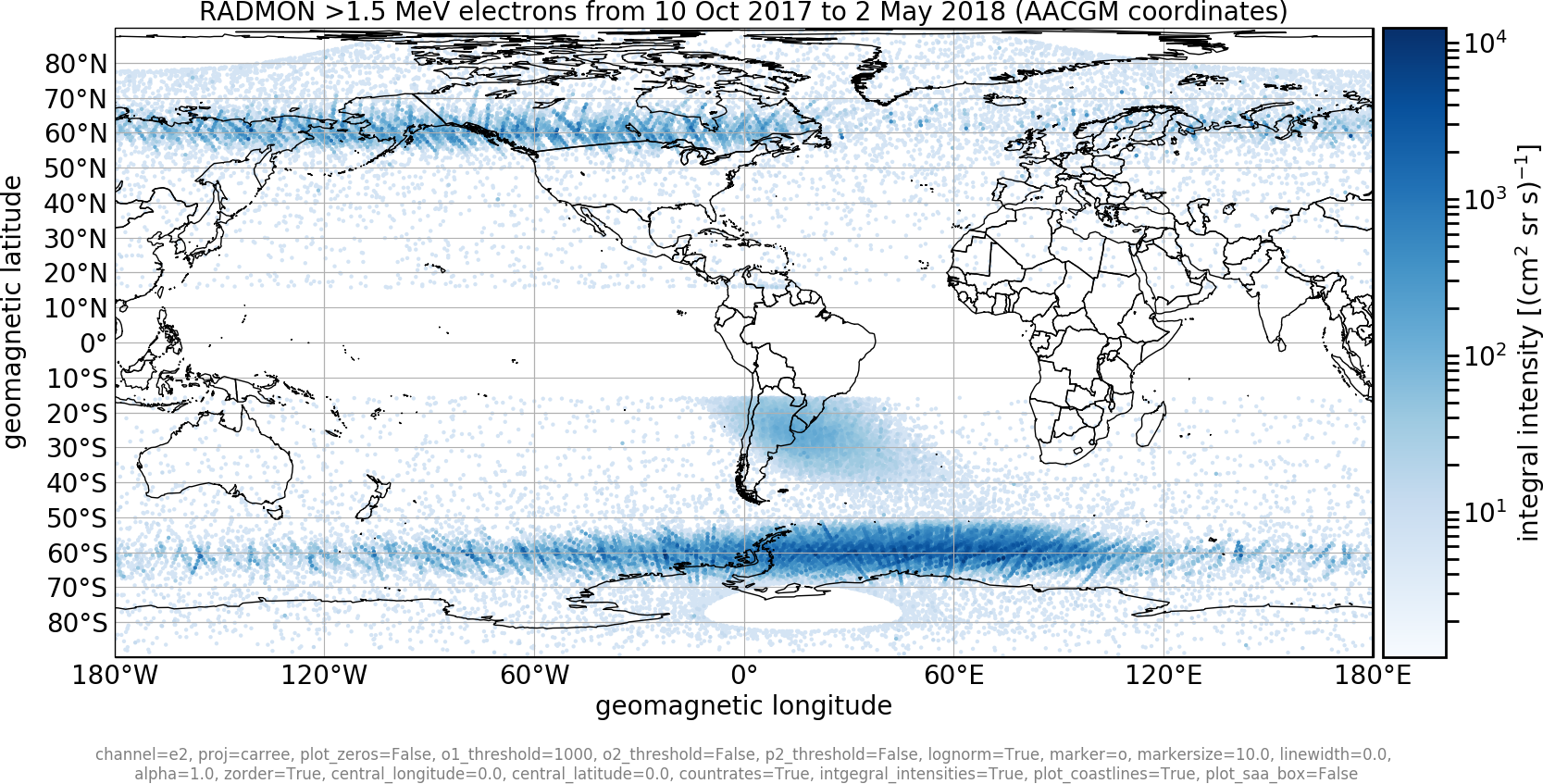}
\caption{Integral intensities of channel e2 (electrons with $>$1.5~MeV) for the main RADMON mission period from 10 Oct 2017 to 2 May 2018 (omitting zero counts) in altitude adjusted corrected geomagnetic (AACGM) coordinates (note that the highest z values are plotted 'on top' of the lower values). 
}
\label{fig:e2_aacgmv2_countorder_blue_o1_lt_1000_bigger_crop}
\end{figure}

\begin{figure}
\includegraphics[width=\columnwidth]{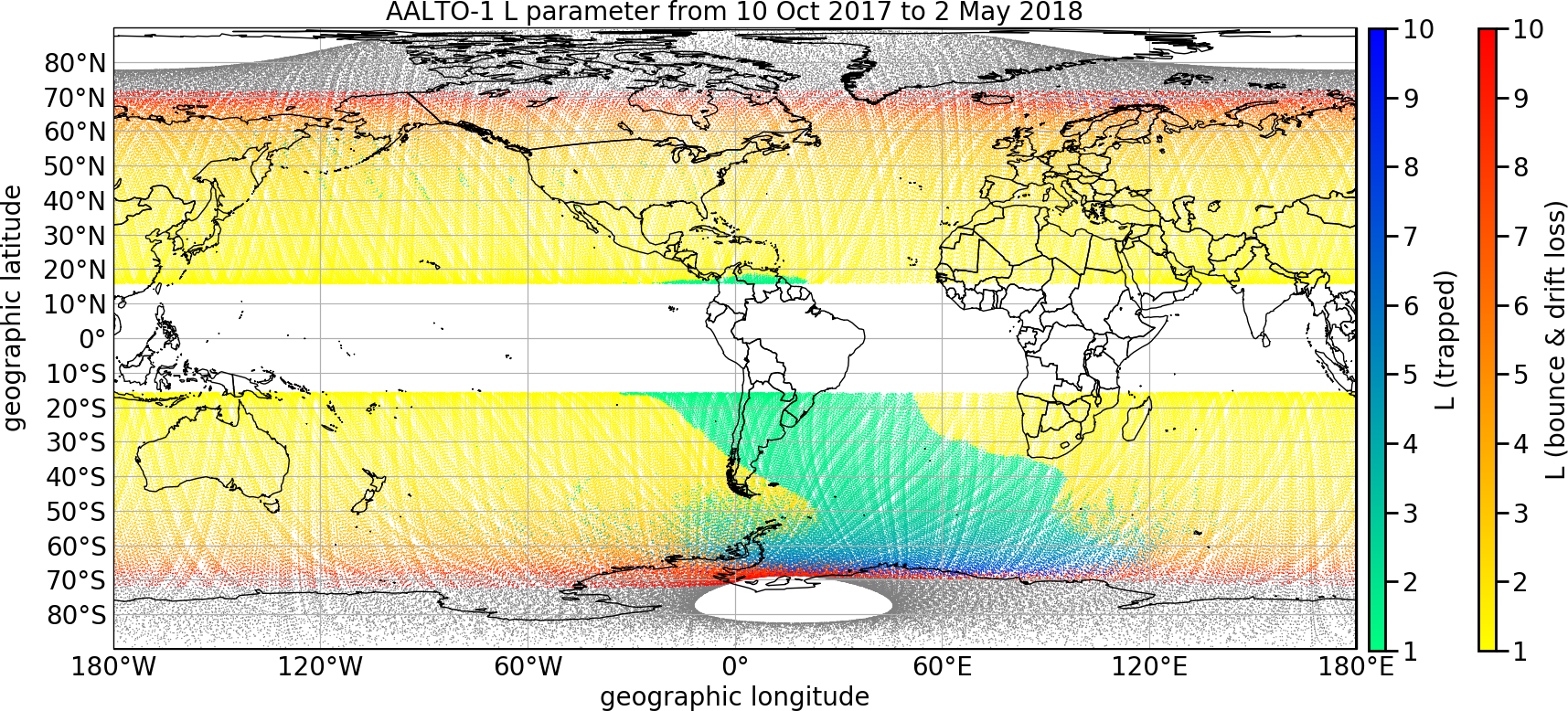}
\caption{Calculated $L$ parameter for Aalto-1 orbits for the main RADMON mission period from 10 Oct 2017 to 2 May 2018 in altitude adjusted corrected geomagnetic (AACGM) coordinates. The two different colorbars for the $L$ values correspond to the trapping conditions of locally mirroring particles: green-to-blue colors mark trapped particles on closed drift shells, yellow-to-red colors particles for which the mirror point in the southern hemisphere is in the loss cone or for which the drift shell hits Earth in the SAA, respectively. (Note that in some parts the trapped data points are plotted 'on top' of the loss data points.)
}
\label{fig:plot_geo_L_RADMON_trapped_ontop}
\end{figure}

Analogous to Fig.~\ref{fig:2017-12-02_plasma_crop}, in Fig.~\ref{fig:e2_countorder_blue_o1_lt_1000_bigger_crop} the $>$1.5~MeV electron intensities of channel e2 are presented for geographical coordinates over the whole mission period. For better visibility, all measurements with an intensity of zero were neglected, and the highest measurement values have been plotted 'on top' of the lower ones. 
With this approach, the latitudes close to the poles can be identified as the regions with the highest detected intensities (sometimes called polar "horns").
For further analysis and in order to allow for a better comparison with other missions at different orbits, we move from the standard geographical coordinate system to altitude adjusted corrected geomagnetic (AACGM) coordinates. See \citet{Shepherd-2014} and reference therein for a detailed description of the AACGM system, and \citet{Burrell-et-al-2018} for the corresponding Python library used for conversion. Figure~\ref{fig:e2_aacgmv2_countorder_blue_o1_lt_1000_bigger_crop} shows the data from Fig.~\ref{fig:e2_countorder_blue_o1_lt_1000_bigger_crop} converted to AACGM coordinates. The overall picture stays the same but the polar "horns" are accurately aligned with latitude in the new AACGM coordinate system.
In this high latitude regions the outer radiation belt reaches down to lower altitudes partly allowing the detection of the trapped particle population. 
This is indicated by Fig.~\ref{fig:plot_geo_L_RADMON_trapped_ontop} which shows the calculated $L$ parameter \citep{McIlwain-1961} for Aalto-1 orbits for the main RADMON mission period from 10 Oct 2017 to 2 May 2018 in AACGM coordinates. 
$L$ values have been calculated with \textit{SpacePy 0.1.6} \citep{Morley-etal-2011} using the \textit{IRBEM 4.3} library \citep{Boscher-et-al} with \citet{Olson-Pfitzer-1974} static model for quiet conditions as external and \textit{IGRF-12} \citep{Thebault-et-al-2015} as internal magnetic field.
The two different colorbars for the $L$ values correspond to the trapping conditions of locally mirroring particles: green-to-blue colors mark trapped particles on closed drift shells, yellow-to-red colors particles for which the mirror point in the southern hemisphere is in the loss cone or for which the drift shell hits Earth in the SAA, respectively.
Note that these trapping conditions correspond to locally mirroring particles. Thus, it is a necessary condition for particles to be trapped. But the trapping region gets smaller and smaller if we allow the local pitch angle to differ from 90$^\circ$.
The gap in the northern "horn" between North America and Europe can be attributed to the tilted Earth magnetic field which causes the mirror points of the trapped particles to be elevated in the northern hemisphere. 
These particles are lost to the atmosphere at the equivalent southern latitudes, visible by the higher intensities in the region south of the SAA. This overall picture has recently also been observed by e.~g. the PROBA-V mission \citep{Pierrard-etal-2014} at a slightly higher altitude of 820 km.
It needs to be pointed out that the apparently high electron intensities detected by RADMON in the SAA are not really electrons of the observed energy but mainly higher energy protons which contaminate and dominate the RADMON electron channels in this region. 
The contaminating proton energies can be estimated as $>$100~MeV (for electron channel e2), $>$80~MeV (e3), $>$70~MeV (e4), and $>$50~MeV (e5).
More details on this instrumental effect can be found in \citet{Oleynik-etal-2019}.%
The combined intensity of all protons detected by RADMON over the whole mission period is presented in AACGM coordinates in Fig.~\ref{fig:p-all_aacgmv2_countorder_red_o1_lt_1000_bigger_crop}. Note that the empty region around the equator originates from the way the AACGM coordinates are derived. These low AACGM latitudes ($<$16.4$^\circ$) are not accessible for the Aalto-1 altitude \citep[cf.][]{Shepherd-2014}.

\section{Observation of magnetospheric dynamics}
\label{sect:magnetospheric-dynamics}
\begin{figure}
\includegraphics[width=\columnwidth]{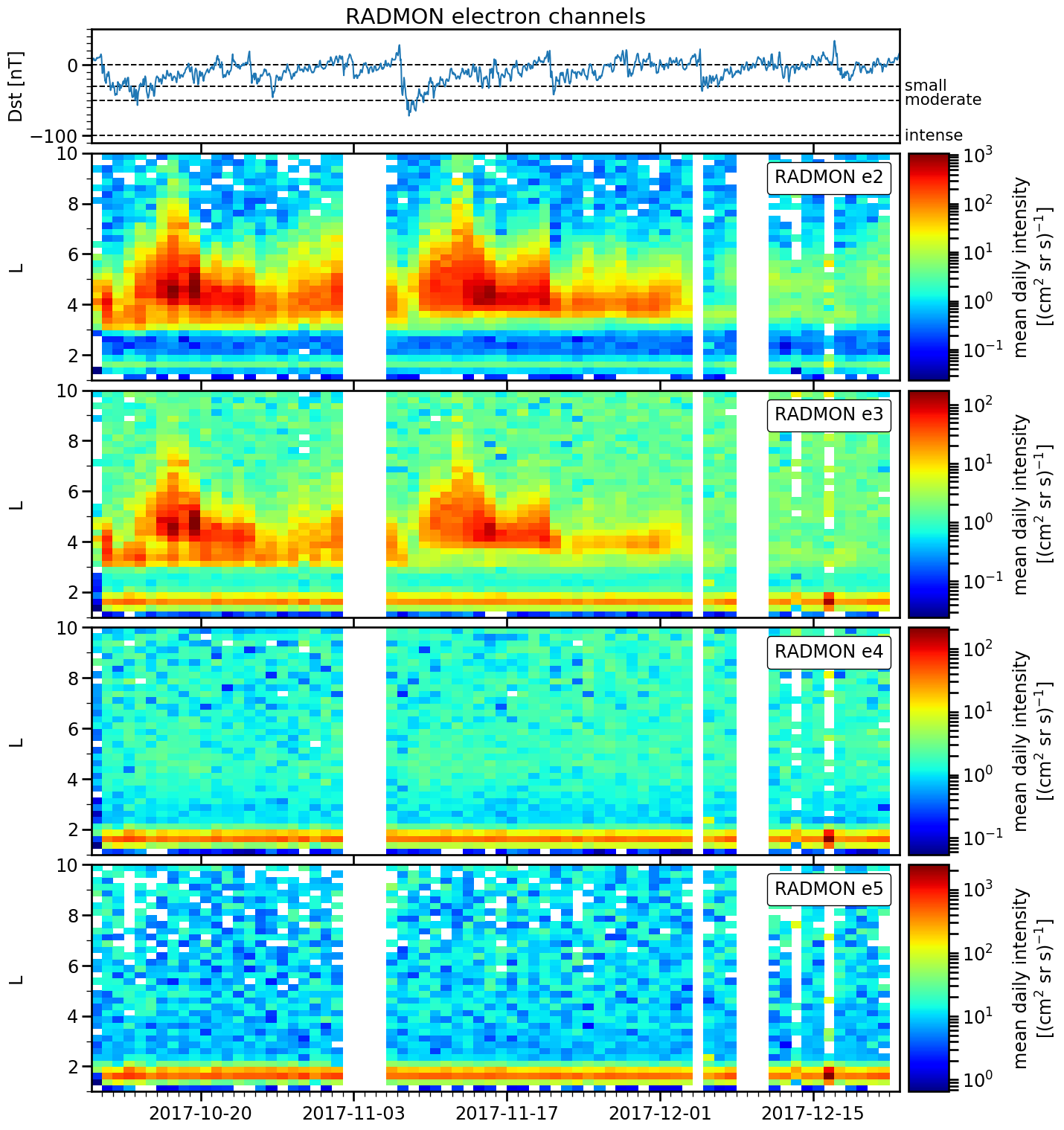}
\caption{Time series of Dst index and histograms of integral intensities obtained by the different RADMON electron channels from 10 Oct 2017 to 21 Dec 2017 with respect to $L$ parameter (z-axis gives color-coded arithmetic daily mean of intensity per bin -- note that the color scale is different for all panels in order to enhance the details of all channels which have different sensitivities).}
\label{fig:lm_int_series_hist_radmon_e}
\end{figure}

\begin{figure}
\includegraphics[width=\columnwidth]{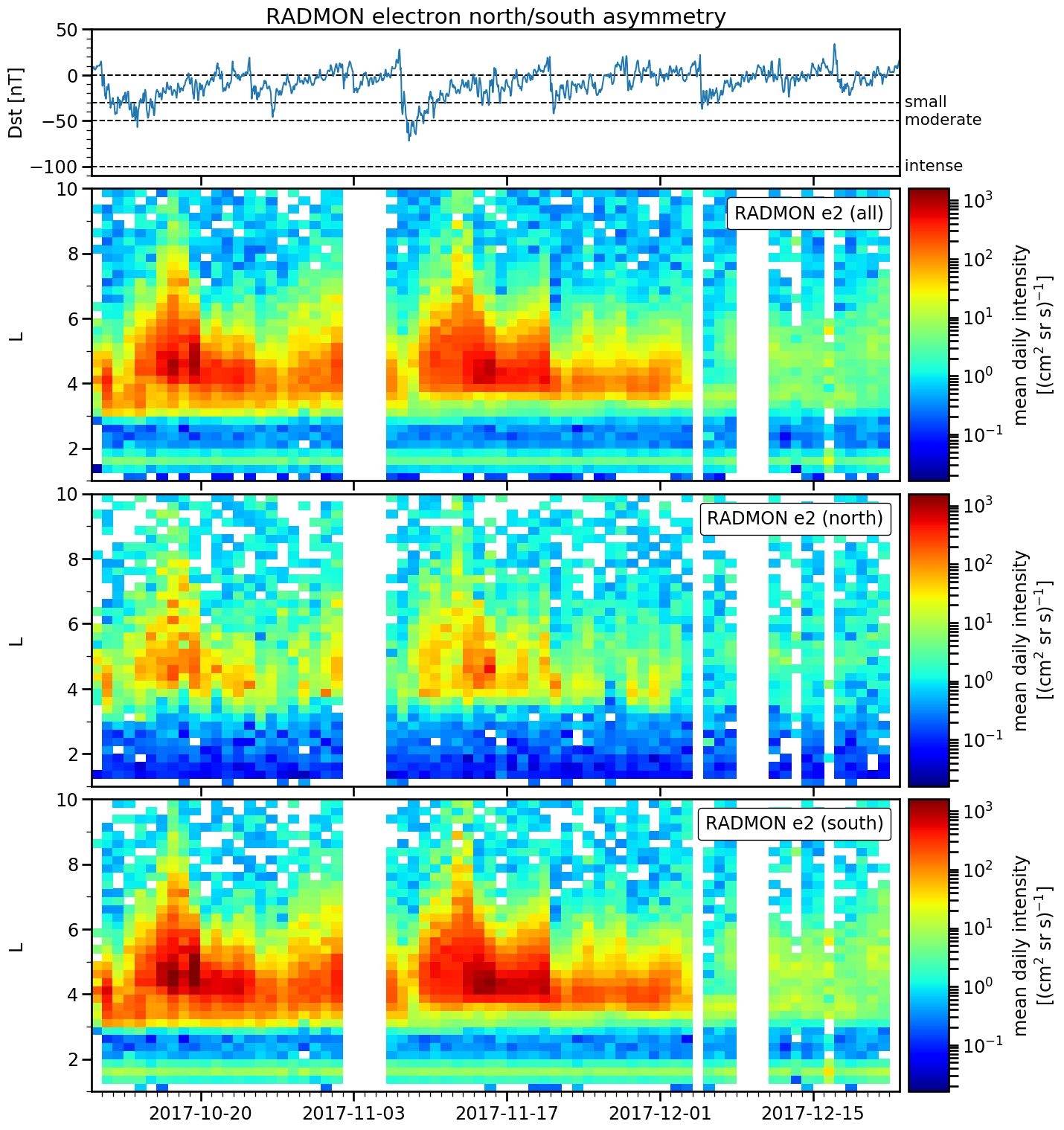}
\caption{Time series of Dst index and histograms of integral intensities obtained by RADMON electron channel e2 ($>$1.5~MeV) from 10 Oct 2017 to 21 Dec 2017 with respect to $L$ parameter 
for all coordinates (top, same as third panel of Fig.~\ref{fig:lm_int_series_hist_radmon_e}) and for the northern (middle) and southern hemisphere (bottom), respectively (z-axis gives color-coded arithmetic daily mean of intensities per bin -- note that the color scale is the same for all panels).
}
\label{fig:lm_int_series_hist_radmon_ns}
\end{figure}

To illustrate the response of RADMON measurements to magnetospheric dynamics, in Fig.~\ref{fig:lm_int_series_hist_radmon_e} the daily electron intensities with respect to the $L$ parameter (as calculated in Sect.~\ref{sect:RADMON-data}) are shown as time series over the main mission phase (green shading in Fig.~\ref{fig:hist_inttime_channels}) for the different instrument channels. 
Above the RADMON electron observations the Dst index obtained by OMNIWeb\footnote{\url{https://omniweb.gsfc.nasa.gov}} is shown as a measure of the geomagnetic activity, with horizontal lines indicating small, moderate and intense storm levels. From Fig.~\ref{fig:lm_int_series_hist_radmon_e} it is apparent that only the RADMON electron channels e2 and e3 depict the dynamic effects of the magnetosphere 
(channel e1 not shown here because it contains almost no data). The almost constant band at $L$ values between 1 and 2 is the observed SAA region, which is dominated by higher energy protons. 
This becomes clearer in Fig.~\ref{fig:lm_int_series_hist_radmon_ns}, where the measurements of electron channel e2 are shown for all AACGM latitudes (top), only northern (middle) and only southern hemisphere (bottom), respectively. The northern hemisphere yields overall lower intensities, and does not have the SAA band at $L<2$. Nevertheless, between $L$ values of 4 and 8 it shows the same temporal structures as the observation in the southern hemisphere, which dominates the picture over all latitudes. The $L>3$ region is connected to the outer belt, and shows much more variability with time. Both storms reaching the moderate storm Dst level produce strong enhancements in the outer belt electrons, while the small storms seem rather to lead to depletion of the outer belt from the most energetic electrons.

In the Appendix, the time series of proton measurements with respect to $L$ value is presented in Fig.~\ref{fig:lm_int_series_hist_radmon_p_i}. 

\section{Comparison with PROBA-V/EPT} 
In order to further verify the RADMON measurements in the magnetosphere, a comparison with observations of the Energetic Particle Telescope (EPT) onboard the ESA minisatellite (volume $<$1m$^3$) PROBA-V (PRoject for OnBoard Autonomy–Vegetation) have been carried out. PROBA-V was launched on 7 May 2013 into a polar LEO with 98.7$^\circ$ inclination and an altitude of 820 km. The EPT is a modular ionizing particle spectrometer consisting of a low energy section made up of two silicon detectors (using the $\Delta$E-E method), and a high energy section. The latter uses ten pairs of absorber material and attached silicon sensor to generate a digital signal with the only information which sensors have been triggered (i.e., yielding a given threshold). With this setup, the EPT is able to detect electrons from 0.5-20~MeV, protons from 9.5-300~MeV and He-ions from 38-1200~MeV \citep{Cyamukungu-etal-2014}.

\subsection{Magnetospheric dynamics}
\begin{figure}
\includegraphics[width=\columnwidth]{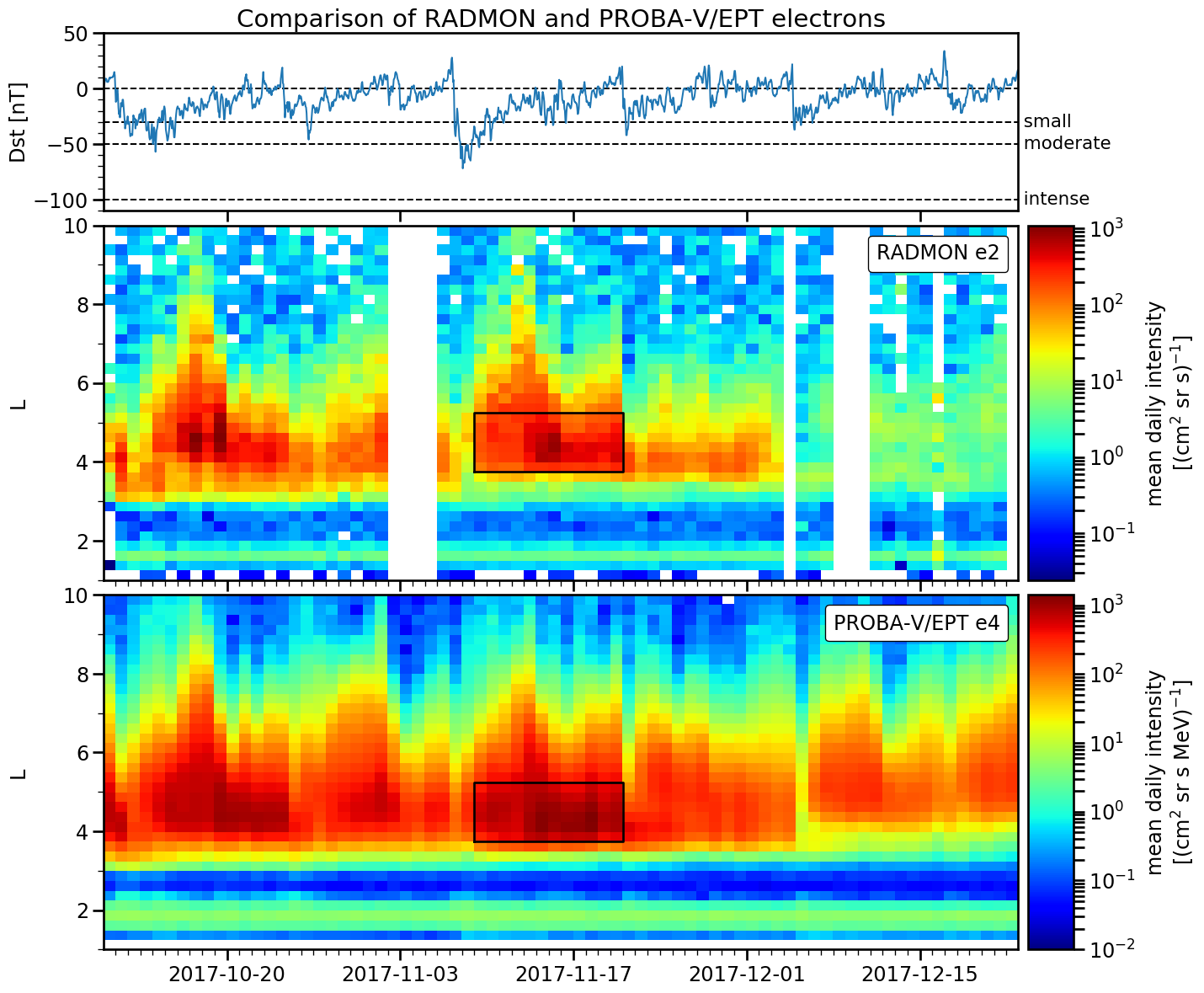}
\caption{Time series of Dst index and histograms of RADMON electron  ($>$1.5~MeV) integral intensities and PROBA-V/EPT electron (1.7~MeV) intensities, respectively, with respect to $L$ parameter from 10 Oct 2017 to 21 Dec 2017 (z-axis gives color-coded arithmetic mean of intensities, respectively, per bin). The boxes indicate the time and $L$ value range used to obtain the spectra in Fig.~\ref{fig:spectrum_nov_L_double_powerlaw}.
}
\label{fig:lm_int_series_hist_comb}
\end{figure}
In Fig.~\ref{fig:lm_int_series_hist_comb} the time series of RADMON e2 electron integral intensities ($>$1.5~MeV) with respect to $L$ value (same as top panels of Figs.~\ref{fig:lm_int_series_hist_radmon_e} and \ref{fig:lm_int_series_hist_radmon_ns}) is shown next to PROBA-V/EPT differential intensity measurements of electrons in the energy range 1--2.4~MeV. Although PROBA-V is at a higher altitude, the overall picture looks quite similar (note that while at comparable energies, the two measurements are in different units). Especially the outer belt dynamics ($L$ values between 3 and 8) show the same temporal structures in the RADMON measurements, although slightly more coarse. In December 2017, however, RADMON e2 seems to be measuring lower intensities compared to the PROBA-V/EPT observations. This seems to be a spectral effect because looking at lower (e3; 0.9 MeV) and higher (e5; 5.2 MeV) EPT energies (Figs.~\ref{fig:lm_int_series_hist_comb_e3} and \ref{fig:lm_int_series_hist_comb_e5}, respectively) shows that EPT sees a clear spectral softening in the beginning of December, which would explain why the integral RADMON e2 channel would suffer a greater loss of signal than the differential EPT e4 channel.
The spectral softening from November to December 2017 is clearly seen in RADMON data as well, with no signal in e4 ($>$6.2~MeV) over the background and even e3 channel ($>$3.1~MeV) background-dominated (cf.~Fig.~\ref{fig:lm_int_series_hist_radmon_e}). Because of the high background in these channels, it is not possible to derive a reliable spectral index from RADMON measurement. But the steep spectrum observed by PROBA-V/EPT (see Sect.~\ref{sect:spectral-comparison}) is consistent with the low levels of RADMON integral channels. The remaining reduction of counts in Aalto-1 orbit could be affected by the pitch-angle distribution at PROBA-V. Unfortunately, PROBA-V/EPT measurements do not cover well 
those pitch angles that would be observed by RADMON, i.e., between the loss-cone edge at $\sim$58$^\circ$ and a value of $\sim$69$^\circ$, which is the limit of pitch angles in PROBA-V orbit for particles mirroring at Aalto-1 orbit.\footnote{The numbers are obtained from the dipole approximation at $L=4$, but do not vary by many degrees.} This region of pitch angles could well be affected by pitch-angle diffusion processes and would not necessarily show the same flux as for locally mirroring particles at PROBA-V. In fact, the observed pitch-angle distribution indicates that the region near the center of the pitch-angle range has higher fluxes than the edges of the PROBA-V range.

\subsection{Spectral comparison}
\label{sect:spectral-comparison}
To compare electron energy spectra of RADMON and PROBA-V/EPT, a period with high intensities is chosen in the $L$-dependent intensity time series (boxes in Fig.~\ref{fig:lm_int_series_hist_comb}). During this selected period both instruments are connected to the outer belt electron population, giving higher statistics and less contamination in the RADMON electron channels.
\begin{figure}
\includegraphics[width=\columnwidth]{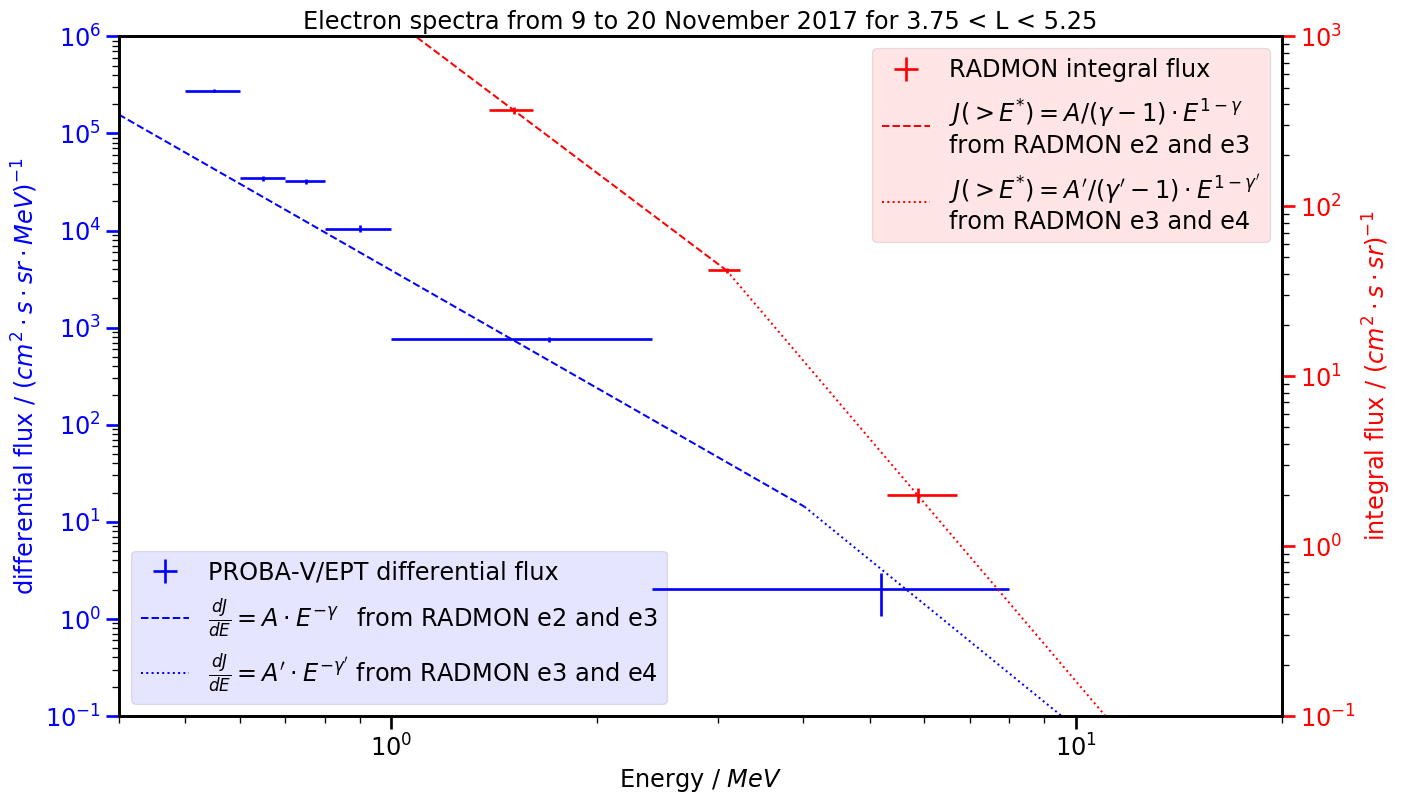}
\caption{Electron spectra from 9 to 20 Nov 2017 and for $3.75 < L < 5.25$ (as indicated by boxes in Fig.~\ref{fig:lm_int_series_hist_comb}) measured in integral flux by RADMON (red data points) and in differential flux by PROBA-V/EPT (blue data points), respectively. The dashed and dotted lines represent fits to the RADMON measurements in integral flux units (red) as well as converted to differential flux units (blue), respectively. RADMON channel e4 ($>$6.2~MeV) includes some amount of proton contamination, which is not subtracted here.
}
\label{fig:spectrum_nov_L_double_powerlaw}
\end{figure}
Figure~\ref{fig:spectrum_nov_L_double_powerlaw} presents the spectra over this 12-day interval. Measured integral intensities of RADMON are shown as red data points with uncertainty ranges at their corresponding cutoff energies (omitting lowest and highest channel e1 and e5 because of low statistics and contamination, respectively) on the red (right) y-axis. On the blue (left) y-axis the differential intensities of PROBA-V/EPT are plotted as data points with the corresponding energy bin limits. In order to obtain comparable units, for the RADMON electron channel pairs e2 and e3 as well as for e3 and e4 two power laws for the integral intensities are fitted to the measurements, resulting in the dashed (dotted) red lines for lower (higher) energies, respectively. The integral intensities each are given by the equations
\begin{eqnarray}
J(>\!E) &=& \frac{A}{\gamma-1}\cdot E^{1-\gamma}\\
J(>\!E) &=& \frac{A^\prime}{\gamma^\prime-1}\cdot E^{1-\gamma^\prime}
\end{eqnarray}
for lower and higher energies. From this equations the values of $A$ and $\gamma$ as well as $A^\prime$ and $\gamma^\prime$ can be obtained. The corresponding RADMON electron differential intensities for the two energy regimes can now be calculated by
\begin{eqnarray}
\frac{dJ}{dE} &=& A\cdot E^{-\gamma} \label{eq:powerlaw3}\\
\frac{dJ}{dE} &=& A^\prime\cdot E^{-\gamma^\prime}
\end{eqnarray}
and plotted as dashed (dotted) blue lines for the lower (higher) energy ranges in Fig.~\ref{fig:spectrum_nov_L_double_powerlaw}.
The RADMON differential intensity spectrum is consistent with EPT measurements at the presented lower RADMON energy range, with RADMON detecting slightly lower intensities for the selected time (and $L$ value) range. This extent of EPT intensities by a factor of $\lesssim$2 is consistent with the higher orbit of PROBA-V compared to Aalto-1. 
RADMON is detecting more counts than EPT at the higher energy channel e4.
On the other hand, this channel is more prone to proton contamination, and from Fig.~\ref{fig:lm_int_series_hist_radmon_e} it is already visible that its statistics are less convincing.

\begin{figure}
\centering
\includegraphics[width=\columnwidth]{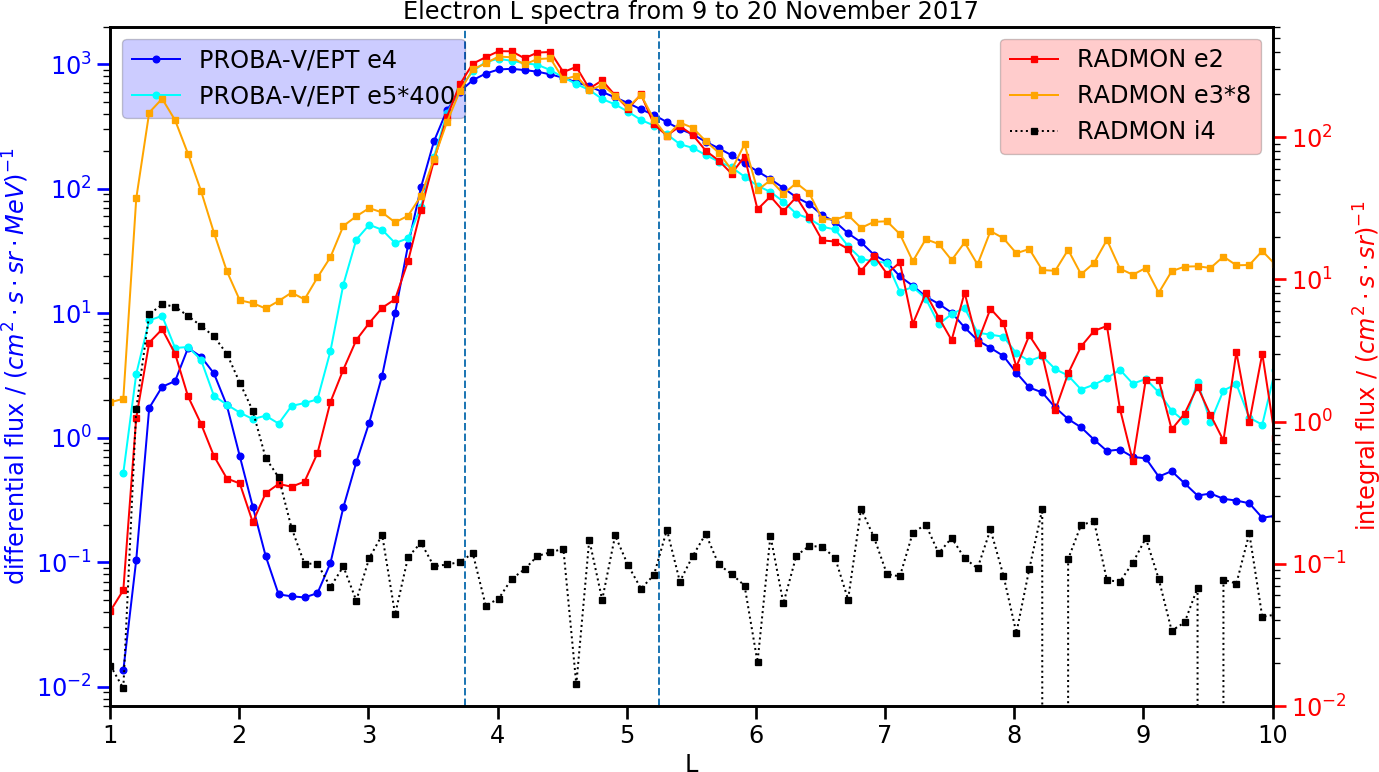}
\caption{$L$ spectra of 1.5~MeV and 1.7~MeV electrons measured by RADMON and PROBA-V/EPT from 9 to 20 Nov 2017 (as indicated by the x-values of the boxes in Fig.~\ref{fig:lm_int_series_hist_comb}). EPT differential intensities (blue legend) refer to the left y-axis, RADMON integral intensities (red legend) to the right y-axis. Plotted by dashed black line is RADMON proton channel i4 ($>$31~MeV). 
Marked by vertical lines is the $L$ range used for the energy spectra calculation in Fig.~\ref{fig:spectrum_nov_L_double_powerlaw}.}
\label{fig:int_l_nov}
\end{figure}
We also analyzed the spectra in the low-intensity period in December, namely 11 to 21 December 2017 at $4.5<L<5.25$. RADMON shows a clear increase over the background only in the channel e2 so the spectral index cannot be determined from those measurements. However, PROBA-V/EPT measurements can be analyzed for the spectral shape. The power-law spectral index around 1~MeV is about $\gamma=7.4$ for this time period. Using that value to convert the corresponding average integral intensity in RADMON e2 to a differential intensity at 1.5~MeV compares very well with the EPT value, being about 60\% of the fitted value at EPT. Thus, we are confident that the clear reduction of RADMON intensities in December 2017 is a real effect and mostly related to the soft spectrum prevailing over the period.

In addition to the electron energy spectra of Fig.~\ref{fig:spectrum_nov_L_double_powerlaw}, in Fig.~\ref{fig:int_l_nov} the $L$ spectra of RADMON and PROBA-V/EPT measurements from 9 to 20 November 2017 at 1.5~MeV (e2) and 1.7~MeV (e4), respectively, are presented. For this purpose, all intensities are assigned correspondingly to the $L$ value at which they were obtained to $L$ bins of width 0.05. For each $L$ bin then the arithmetic mean is calculated. Note that EPT measurements of differential intensities are plotted with respect to the left y-axis, whereas RADMON's integral intensities refer to the right y-axis. RADMON electron channels e2 and e3 are sensitive to high energy protons of $>$100~MeV and $>$80~MeV, respectively. Because of that in addition the RADMON proton channel i4 ($>$31~MeV) is shown.
Except for very low $L$ values $<$2.5 where proton contamination from the inner belt is altering the results, the spectral shapes of RADMON e2 and EPT e4 observations are in good agreement. 
This is especially true for the range $3.75 < L < 5.25$ (marked by vertical lines) for which the energy spectra were calculated and compared. Only for the higher end of analyzed $L$ values the RADMON e2 intensities are varying more and tend to differ from the EPT e4 findings. In this range, as for the low $L$ values, the statistics are worse compared to the intermediate range. 
Around $L$ values of 2.5-3 RADMON e2 shows some kind of bulge in the intensities which is not observed by EPT e4. The reason for this could be consistent with a spectral change in this region because in contrast to EPT the RADMON electron channels are integral channels which are sensitive to higher energies, too. And such a bulge-like intensity enhancement can also be seen in the higher energy channels EPT e5 and RADMON e3.

\section{RADMON data release}
The RADMON data set is made publicly available\footnote{\url{https://zenodo.org/record/3541628} (doi:10.5281/zenodo.3541628)} in energy channels provided in Tab.~\ref{tab:channels}. The electron channels e2 to e4 are the electron channels directly provided by the measurements. The proton channels i1 to i5 are obtained by summing the counts in the measurement channels, i.e., i1 corresponds to protons in channels p1 to p9, i2 to protons in channels p2 to p9, and so on. This allows us to provide the measurements in channels with well-defined responses that are much easier to interpret than the directly measured proton channels that have more irregular responses. The response functions are presented in detail in \citet{Oleynik-etal-2019}.

\begin{table}
\caption{RADMON energy channels included in the data release.\label{tab:channels}}
\begin{centering}
\begin{tabular}{cccc}
    \hline
    Name & Primary species & Contaminating species & Units\\
    \hline
    e2 & $>$1.5 MeV electron & $>$100 MeV proton& (cm$^{2}$ sr s)$^{-1}$\\
    e3 & $>$3.1 MeV electron & $>$80 MeV proton& (cm$^{2}$ sr s)$^{-1}$\\
    e4 & $>$6.0 MeV electron & $>$70 MeV proton& (cm$^{2}$ sr s)$^{-1}$\\
    i1 & $>$10.4 MeV proton & -- & (cm$^{2}$ sr s)$^{-1}$\\
    i2 & $>$18.5 MeV proton & -- & (cm$^{2}$ sr s)$^{-1}$\\
    i3 & $>$23.7 MeV proton & -- & (cm$^{2}$ sr s)$^{-1}$\\
    i4 & $>$29 MeV proton & -- & (cm$^{2}$ sr s)$^{-1}$\\
    i5 & 40--80 MeV proton & -- & (cm$^{2}$ sr s MeV)$^{-1}$\\
    \hline
\end{tabular}
\end{centering}
\end{table}

\section{Summary}
In this paper we have presented the first results of the RADMON instrument on-board the Aalto-1 CubeSat. It has been demonstrated that RADMON -- although very limited in its available resources -- is capable of measuring the integral intensities of electrons above 1.5~MeV and protons above 10~MeV in LEO, reflecting the dynamics of the radiation belts in the magnetosphere. 
A comparison of electron energy spectra obtained by PROBA-V/EPT at a $\sim$300~km higher altitude shows that RADMON's 1.5~MeV measurements are in agreement with the expectations of increasing intensities with height, while at higher energies RADMON and EPT indicate some discrepancies. An additional analysis of $L$ spectra obtained for electrons of 1.5~MeV (RADMON) and 1.7~MeV (EPT), respectively, demonstrates that at intermediate $L$ values between 3.5 and 8, where a connection to the outer belt is prevalent, both missions show the same $L$ dependence for intensities. 
A data set with reliable RADMON measurements has been made available to the scientific community.

\paragraph{Acknowledgements}
This work was performed in the framework of the Finnish Centre of Excellence in Research of Sustainable Space (FORESAIL) funded by the Academy of Finland (grants 312357 and 312358).
The authors are grateful to the PROBA-V/EPT Team at CSR for providing the L1 data, as well as to the Belgian Science Policy Office (Belspo) and the European Space Agency (ESA) for supporting the PROBA-V/EPT project. 
We also gratefully acknowledge the efforts of dozens of students in Aalto University, University of Turku and University of Helsinki for their work in the Aalto-1 satellite and RADMON projects.
HH is supported by the Turku Collegium for Science and Medicine.
Aalto University MIDE is thanked for financial support for building Aalto-1. Aalto University, University of Turku, RUAG, Space Systems Finland, and Nokia sponsored the launch of the satellite.

\appendix
\section{Additional figures}
\begin{figure}
\centering
\includegraphics[width=\columnwidth]{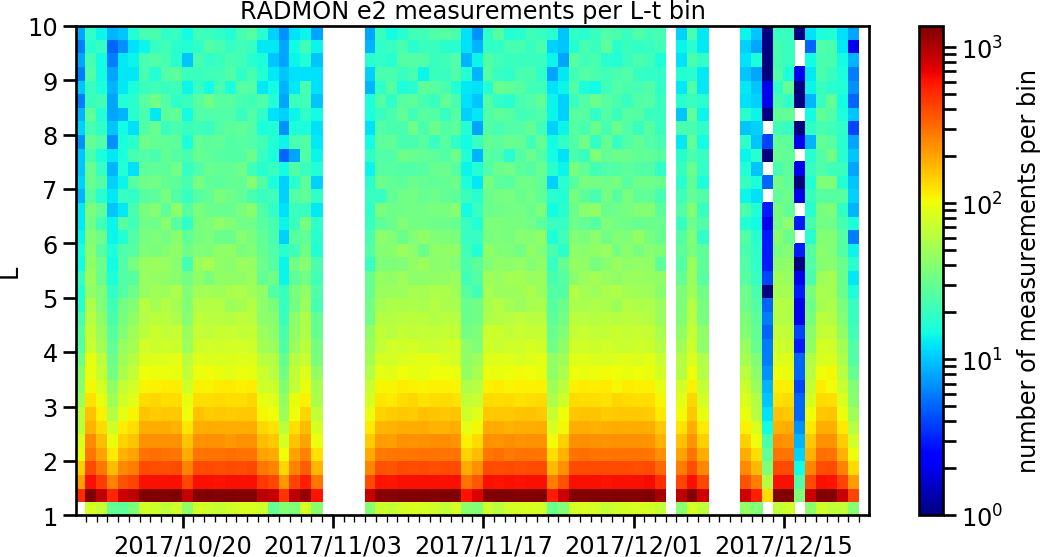}
\caption{Time series of number of RADMON electron ($>$1.5~MeV) intensity measurements per $L$-$t$ bin (with resolution $\Delta L=0.25$ and $\Delta t=1$d) from 10 Oct 2017 to 21 Dec 2017 (z-axis gives color-coded the number of measurements per bin). See Sect.~\ref{sect:RADMON-data} for more details on the $L$ calculation.
}
\label{fig:lm_n_series_hist_radmon_e2_crop}
\end{figure}
\begin{figure}
\centering
\includegraphics[width=\columnwidth]{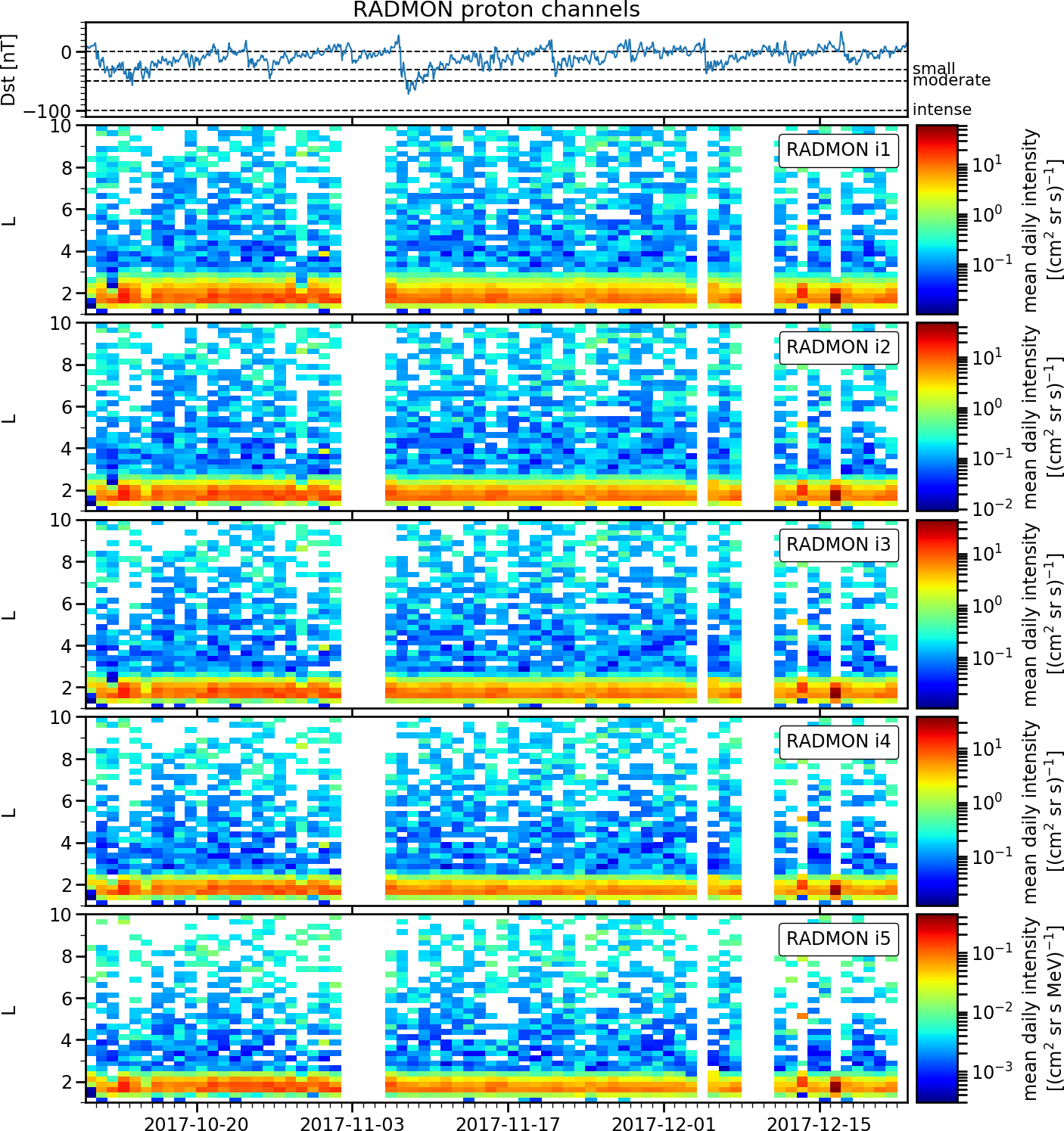}
\caption{Time series of Dst index and histograms of proton intensities with respect to $L$ parameter obtained by the different RADMON proton combination channels (i1 to i5 in Tab.~\ref{tab:channels}) from 10 Oct 2017 to 21 Dec 2017 (z-axis gives color-coded arithmetic mean of count rate per bin). Note that channels i1 to i4 are integral intensities whereas i5 are differential intensities.
}
\label{fig:lm_int_series_hist_radmon_p_i}
\end{figure}
\begin{figure}
\includegraphics[width=\columnwidth]{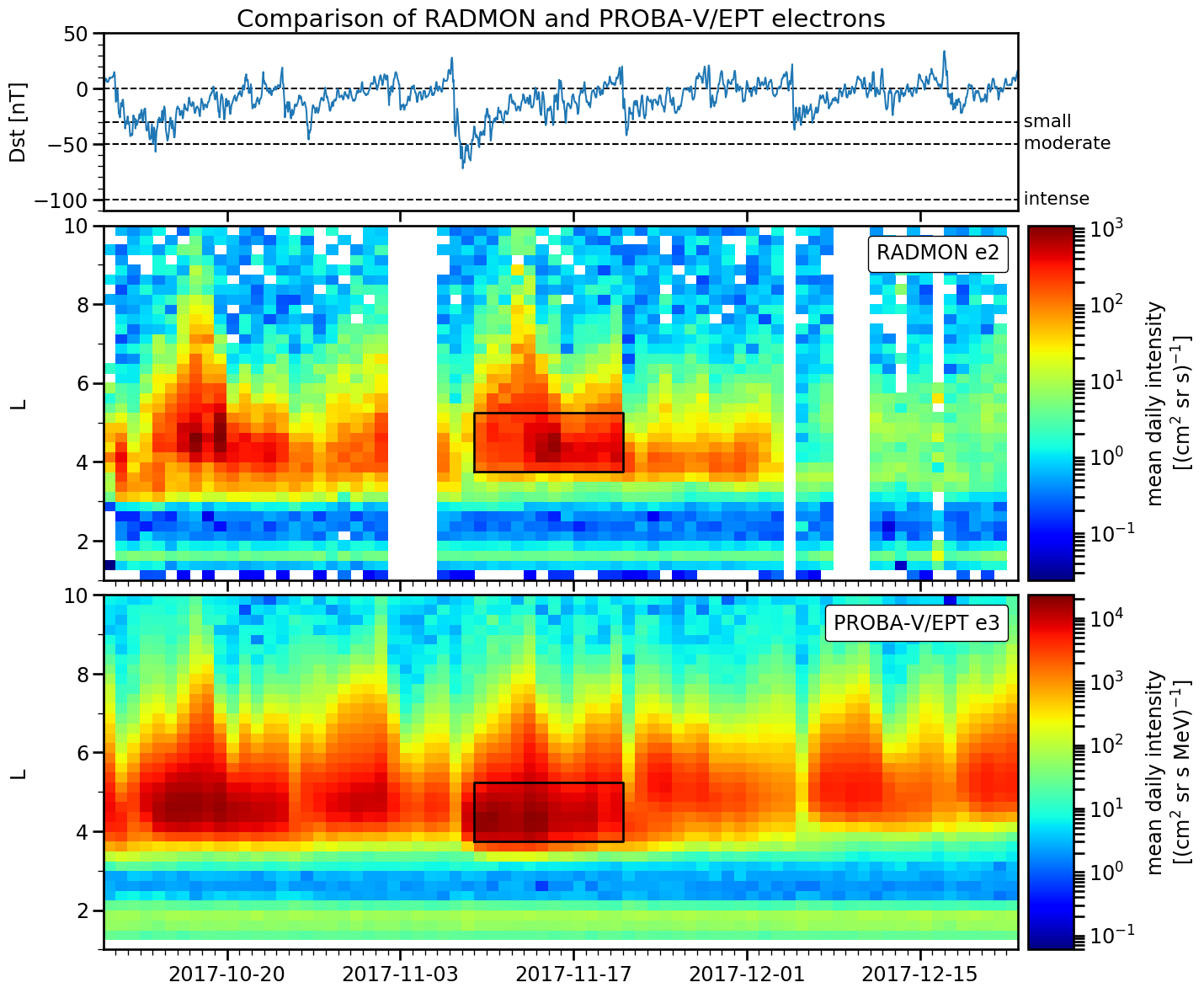}
\caption{Time series of Dst index and histograms of RADMON electron ($>$1.5~MeV) integral intensities and PROBA-V/EPT electron (0.9~MeV) intensities, respectively, with respect to $L$ parameter from 10 Oct 2017 to 21 Dec 2017 (z-axis gives color-coded arithmetic mean of intensities, respectively, per bin). The boxes indicate the time and $L$ value range used to obtain the spectra in Fig.~\ref{fig:spectrum_nov_L_double_powerlaw}.
}
\label{fig:lm_int_series_hist_comb_e3}
\end{figure}
\begin{figure}
\includegraphics[width=\columnwidth]{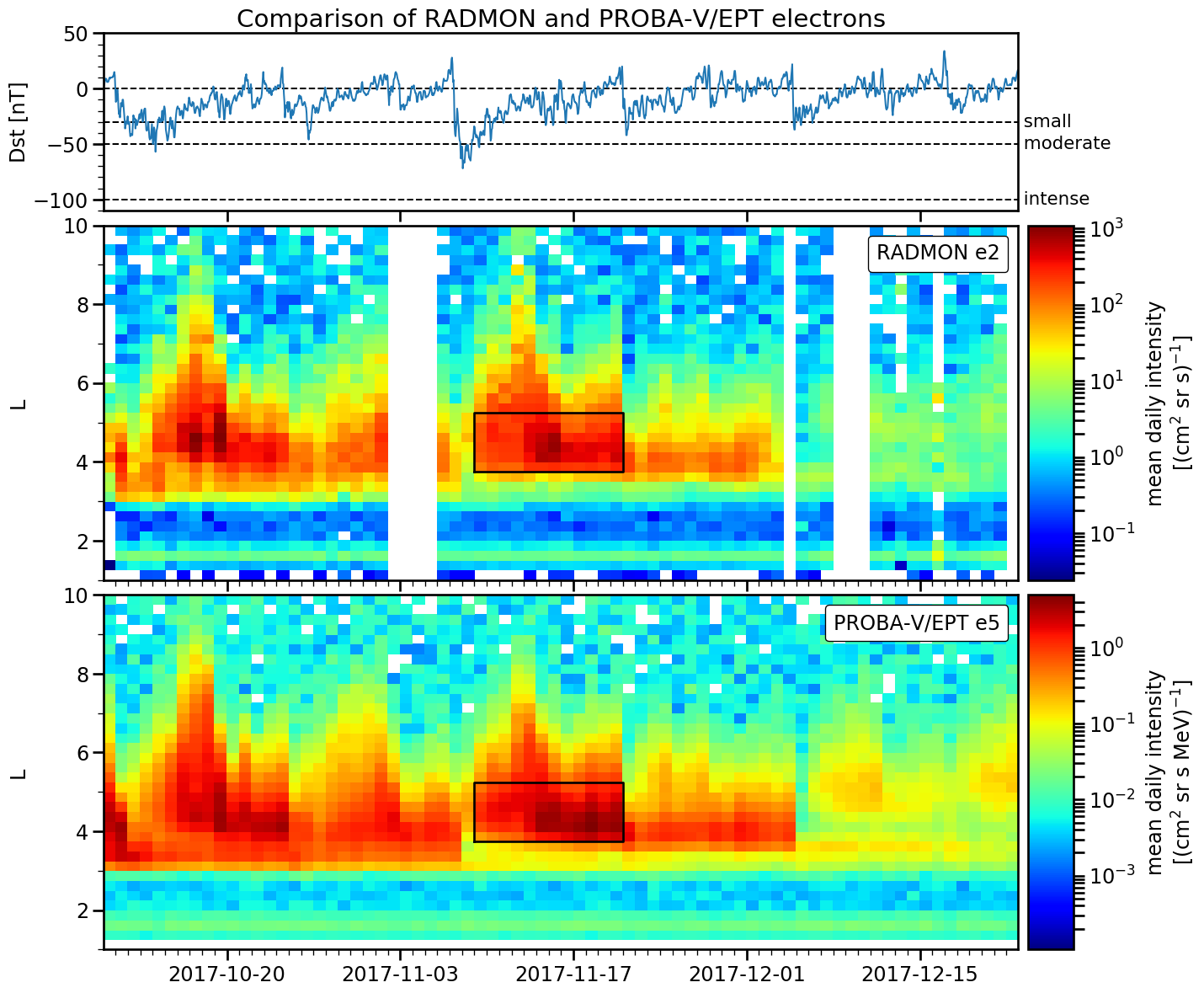}
\caption{Time series of Dst index and histograms of RADMON electron ($>$1.5~MeV) integral intensities and PROBA-V/EPT electron (5.2~MeV) intensities, respectively, with respect to $L$ parameter from 10 Oct 2017 to 21 Dec 2017 (z-axis gives color-coded arithmetic mean of intensities, respectively, per bin). The boxes indicate the time and $L$ value range used to obtain the spectra in Fig.~\ref{fig:spectrum_nov_L_double_powerlaw}.
}
\label{fig:lm_int_series_hist_comb_e5}
\end{figure}
%

\def\aj{AJ}%
\def\actaa{Acta Astron.}%
\def\araa{ARA\&A}%
\def\apj{ApJ}%
\def\apjl{ApJ}%
\def\apjs{ApJS}%
\def\ao{Appl.~Opt.}%
\def\apss{Ap\&SS}%
\def\aap{A\&A}%
\def\aapr{A\&A~Rev.}%
\def\aaps{A\&AS}%
\def\azh{AZh}%
\def\baas{BAAS}%
\def\bac{Bull. astr. Inst. Czechosl.}%
\def\caa{Chinese Astron. Astrophys.}%
\def\cjaa{Chinese J. Astron. Astrophys.}%
\def\icarus{Icarus}%
\def\jcap{J. Cosmology Astropart. Phys.}%
\def\jrasc{JRASC}%
\def\mnras{MNRAS}%
\def\memras{MmRAS}%
\def\na{New A}%
\def\nar{New A Rev.}%
\def\pasa{PASA}%
\def\pra{Phys.~Rev.~A}%
\def\prb{Phys.~Rev.~B}%
\def\prc{Phys.~Rev.~C}%
\def\prd{Phys.~Rev.~D}%
\def\pre{Phys.~Rev.~E}%
\def\prl{Phys.~Rev.~Lett.}%
\def\pasp{PASP}%
\def\pasj{PASJ}%
\def\qjras{QJRAS}%
\def\rmxaa{Rev. Mexicana Astron. Astrofis.}%
\def\skytel{S\&T}%
\def\solphys{Sol.~Phys.}%
\def\sovast{Soviet~Ast.}%
\def\ssr{Space~Sci.~Rev.}%
\def\zap{ZAp}%
\def\nat{Nature}%
\def\iaucirc{IAU~Circ.}%
\def\aplett{Astrophys.~Lett.}%
\def\apspr{Astrophys.~Space~Phys.~Res.}%
\def\bain{Bull.~Astron.~Inst.~Netherlands}%
\def\fcp{Fund.~Cosmic~Phys.}%
\def\gca{Geochim.~Cosmochim.~Acta}%
\def\grl{Geophys.~Res.~Lett.}%
\def\jcp{J.~Chem.~Phys.}%
\def\jgr{J.~Geophys.~Res.}%
\def\jqsrt{J.~Quant.~Spec.~Radiat.~Transf.}%
\def\memsai{Mem.~Soc.~Astron.~Italiana}%
\def\nphysa{Nucl.~Phys.~A}%
\def\physrep{Phys.~Rep.}%
\def\physscr{Phys.~Scr}%
\def\planss{Planet.~Space~Sci.}%
\def\procspie{Proc.~SPIE}%


\bibliography{mybibfile}

\begin{thebibliography}{17}
\expandafter\ifx\csname natexlab\endcsname\relax\def\natexlab#1{#1}\fi
\providecommand{\url}[1]{\texttt{#1}}
\providecommand{\href}[2]{#2}
\providecommand{\path}[1]{#1}
\providecommand{\DOIprefix}{doi:}
\providecommand{\ArXivprefix}{arXiv:}
\providecommand{\URLprefix}{URL: }
\providecommand{\Pubmedprefix}{pmid:}
\providecommand{\doi}[1]{\href{http://dx.doi.org/#1}{\path{#1}}}
\providecommand{\Pubmed}[1]{\href{pmid:#1}{\path{#1}}}
\providecommand{\bibinfo}[2]{#2}
\ifx\xfnm\relax \def\xfnm[#1]{\unskip,\space#1}\fi
\bibitem[{{Boscher} et~al.(2004-2008){Boscher}, {Bourdarie}, {O'Brien} and
  {Guild}}]{Boscher-et-al}
\bibinfo{author}{{Boscher}, D.}, \bibinfo{author}{{Bourdarie}, S.},
  \bibinfo{author}{{O'Brien}, P.}, \bibinfo{author}{{Guild}, T.},
  \bibinfo{year}{2004-2008}.
\newblock \bibinfo{title}{{IRBEM library V4.3}}.
\newblock \URLprefix \url{https://sourceforge.net/projects/irbem/}.
\bibitem[{{Burrell} et~al.(2018){Burrell}, {van der Meeren} and
  {Laundal}}]{Burrell-et-al-2018}
\bibinfo{author}{{Burrell}, A.}, \bibinfo{author}{{van der Meeren}, C.},
  \bibinfo{author}{{Laundal}, K.M.}, \bibinfo{year}{2018}.
\newblock \bibinfo{title}{{Python library for AACGM-v2 magnetic coordinates}}.
\newblock \DOIprefix\doi{10.5281/zenodo.1469697}.
\bibitem[{Cyamukungu et~al.(2014)Cyamukungu, Benck, Borisov, Gregoire, Cabrera,
  Bonnet, Desoete, Preudrhomme, Semaille, Creve, Saedeleer, Ilsen, {De Busser},
  Pierrard, Bonnewijn, Maes, Ransbeeck, Neefs, Lemaire, Valtonen, Punkkinen,
  Anciaux, Litefti, Brun, Pauwels, Quevrin, Moreau, Helderweirt, Hajdas and
  Nieminen}]{Cyamukungu-etal-2014}
\bibinfo{author}{Cyamukungu, M.}, \bibinfo{author}{Benck, S.},
  \bibinfo{author}{Borisov, S.}, \bibinfo{author}{Gregoire, G.},
  \bibinfo{author}{Cabrera, J.}, \bibinfo{author}{Bonnet, J.L.},
  \bibinfo{author}{Desoete, B.}, \bibinfo{author}{Preudrhomme, F.},
  \bibinfo{author}{Semaille, C.}, \bibinfo{author}{Creve, G.},
  \bibinfo{author}{Saedeleer, J.D.}, \bibinfo{author}{Ilsen, S.},
  \bibinfo{author}{{De Busser}, L.}, \bibinfo{author}{Pierrard, V.},
  \bibinfo{author}{Bonnewijn, S.}, \bibinfo{author}{Maes, J.},
  \bibinfo{author}{Ransbeeck, E.V.}, \bibinfo{author}{Neefs, E.},
  \bibinfo{author}{Lemaire, J.}, \bibinfo{author}{Valtonen, E.},
  \bibinfo{author}{Punkkinen, R.}, \bibinfo{author}{Anciaux, M.},
  \bibinfo{author}{Litefti, K.}, \bibinfo{author}{Brun, N.},
  \bibinfo{author}{Pauwels, D.}, \bibinfo{author}{Quevrin, C.},
  \bibinfo{author}{Moreau, D.}, \bibinfo{author}{Helderweirt, A.},
  \bibinfo{author}{Hajdas, W.}, \bibinfo{author}{Nieminen, P.},
  \bibinfo{year}{2014}.
\newblock \bibinfo{title}{{The Energetic Particle Telescope (EPT) on Board
  PROBA-V: Description of a New Science-Class Instrument for Particle Detection
  in Space}}.
\newblock \bibinfo{journal}{IEEE Transactions on Nuclear Science}
  \bibinfo{volume}{61}, \bibinfo{pages}{3667--3681}.
\newblock \DOIprefix\doi{10.1109/TNS.2014.2361955}.
\bibitem[{Janhunen(2010)}]{Janhunen-2010}
\bibinfo{author}{Janhunen, P.}, \bibinfo{year}{2010}.
\newblock \bibinfo{title}{Electrostatic plasma brake for deorbiting a
  satellite}.
\newblock \bibinfo{journal}{Journal of Propulsion and Power}
  \bibinfo{volume}{26}, \bibinfo{pages}{370--372}.
\newblock \DOIprefix\doi{10.2514/1.47537}.
\bibitem[{{Kestil{\"a}} et~al.(2013){Kestil{\"a}}, {Tikka}, {Peitso},
  {Rantanen}, {N{\"a}sil{\"a}}, {Nordling}, {Saari}, {Vainio}, {Janhunen},
  {Praks} and {Hallikainen}}]{Kestila-etal-2013}
\bibinfo{author}{{Kestil{\"a}}, A.}, \bibinfo{author}{{Tikka}, T.},
  \bibinfo{author}{{Peitso}, P.}, \bibinfo{author}{{Rantanen}, J.},
  \bibinfo{author}{{N{\"a}sil{\"a}}, A.}, \bibinfo{author}{{Nordling}, K.},
  \bibinfo{author}{{Saari}, H.}, \bibinfo{author}{{Vainio}, R.},
  \bibinfo{author}{{Janhunen}, P.}, \bibinfo{author}{{Praks}, J.},
  \bibinfo{author}{{Hallikainen}, M.}, \bibinfo{year}{2013}.
\newblock \bibinfo{title}{{Aalto-1 nanosatellite - technical description and
  mission objectives}}.
\newblock \bibinfo{journal}{Geoscientific Instrumentation, Methods and Data
  Systems} \bibinfo{volume}{2}, \bibinfo{pages}{121--130}.
\newblock \DOIprefix\doi{10.5194/gi-2-121-2013}.
\bibitem[{{Li} et~al.(2013){Li}, {Schiller}, {Blum}, {Califf}, {Zhao}, {Tu},
  {Turner}, {Gerhardt}, {Palo} and {Kanekal}}]{Li-etal-2013}
\bibinfo{author}{{Li}, X.}, \bibinfo{author}{{Schiller}, Q.},
  \bibinfo{author}{{Blum}, L.}, \bibinfo{author}{{Califf}, S.},
  \bibinfo{author}{{Zhao}, H.}, \bibinfo{author}{{Tu}, W.},
  \bibinfo{author}{{Turner}, D.L.}, \bibinfo{author}{{Gerhardt}, D.},
  \bibinfo{author}{{Palo}, S.}, \bibinfo{author}{{Kanekal}, S.},
  \bibinfo{year}{2013}.
\newblock \bibinfo{title}{{First results from CSSWE CubeSat: Characteristics of
  relativistic electrons in the near-Earth environment during the October 2012
  magnetic storms}}.
\newblock \bibinfo{journal}{J. Geophys. Res. (Space Physics)}
  \bibinfo{volume}{118}, \bibinfo{pages}{6489--6499}.
\newblock \DOIprefix\doi{10.1002/2013JA019342}.
\bibitem[{{Li} et~al.(2017){Li}, {Selesnick}, {Schiller}, {Zhang}, {Zhao},
  {Baker} and {Temerin}}]{Li-etal-2017}
\bibinfo{author}{{Li}, X.}, \bibinfo{author}{{Selesnick}, R.},
  \bibinfo{author}{{Schiller}, Q.}, \bibinfo{author}{{Zhang}, K.},
  \bibinfo{author}{{Zhao}, H.}, \bibinfo{author}{{Baker}, D.N.},
  \bibinfo{author}{{Temerin}, M.A.}, \bibinfo{year}{2017}.
\newblock \bibinfo{title}{{Measurement of electrons from albedo neutron decay
  and neutron density in near-Earth space}}.
\newblock \bibinfo{journal}{\nat} \bibinfo{volume}{552},
  \bibinfo{pages}{382--385}.
\newblock \DOIprefix\doi{10.1038/nature24642}.
\bibitem[{{McIlwain}(1961)}]{McIlwain-1961}
\bibinfo{author}{{McIlwain}, C.E.}, \bibinfo{year}{1961}.
\newblock \bibinfo{title}{{Coordinates for Mapping the Distribution of
  Magnetically Trapped Particles}}.
\newblock \bibinfo{journal}{\jgr} \bibinfo{volume}{66},
  \bibinfo{pages}{3681--3691}.
\newblock \DOIprefix\doi{10.1029/JZ066i011p03681}.
\bibitem[{Morley et~al.(2010)Morley, Welling, Koller, Larsen, Henderson and
  Niehof}]{Morley-etal-2011}
\bibinfo{author}{Morley, S.K.}, \bibinfo{author}{Welling, D.T.},
  \bibinfo{author}{Koller, J.}, \bibinfo{author}{Larsen, B.A.},
  \bibinfo{author}{Henderson, M.G.}, \bibinfo{author}{Niehof, J.},
  \bibinfo{year}{2010}.
\newblock \bibinfo{title}{{SpacePy - A Python-based Library of Tools for the
  Space Sciences}}, in: \bibinfo{editor}{van~der Walt, S.},
  \bibinfo{editor}{Millman, J.} (Eds.), \bibinfo{booktitle}{Proceedings of the
  9th Python in Science Conference}, pp. \bibinfo{pages}{39 -- 45}.
\bibitem[{{Oleynik} et~al.(2019){Oleynik}, {Vainio}, {Punkkinen}, {Dudnik},
  {Gieseler}, {Hedman}, {Hietala}, {H{\ae}ggstr\"om}, {Niemel\"a}, {Peltonen},
  {Praks}, {Punkkinen}, {S\"antti} and {Valtonen}}]{Oleynik-etal-2019}
\bibinfo{author}{{Oleynik}, P.}, \bibinfo{author}{{Vainio}, R.},
  \bibinfo{author}{{Punkkinen}, A.}, \bibinfo{author}{{Dudnik}, O.},
  \bibinfo{author}{{Gieseler}, J.}, \bibinfo{author}{{Hedman}, H.},
  \bibinfo{author}{{Hietala}, H.}, \bibinfo{author}{{H{\ae}ggstr\"om}, E.},
  \bibinfo{author}{{Niemel\"a}, P.}, \bibinfo{author}{{Peltonen}, J.},
  \bibinfo{author}{{Praks}, J.}, \bibinfo{author}{{Punkkinen}, R.},
  \bibinfo{author}{{S\"antti}, T.}, \bibinfo{author}{{Valtonen}, E.},
  \bibinfo{year}{2019}.
\newblock \bibinfo{title}{{Calibration of RADMON Radiation Monitor Onboard
  Aalto-1 CubeSat}}.
\newblock \bibinfo{journal}{Advances in Space Research (accepted)} .
\bibitem[{{Olson} and {Pfitzer}(1974)}]{Olson-Pfitzer-1974}
\bibinfo{author}{{Olson}, W.P.}, \bibinfo{author}{{Pfitzer}, K.A.},
  \bibinfo{year}{1974}.
\newblock \bibinfo{title}{{A quantitative model of the magnetospheric magnetic
  field}}.
\newblock \bibinfo{journal}{\jgr} \bibinfo{volume}{79},
  \bibinfo{pages}{3739--3748}.
\newblock \DOIprefix\doi{10.1029/JA079i025p03739}.
\bibitem[{Peltonen et~al.(2014)Peltonen, Hedman, Ilmanen, Lindroos, Maattanen,
  Pesonen, Punkkinen, Punkkinen, Vainio, Valtonen, Santti, Pentik{\"{a}}inen
  and Haeggstrom}]{Peltonen-etal-2014}
\bibinfo{author}{Peltonen, J.}, \bibinfo{author}{Hedman, H.P.},
  \bibinfo{author}{Ilmanen, A.}, \bibinfo{author}{Lindroos, M.},
  \bibinfo{author}{Maattanen, M.}, \bibinfo{author}{Pesonen, J.},
  \bibinfo{author}{Punkkinen, R.}, \bibinfo{author}{Punkkinen, A.},
  \bibinfo{author}{Vainio, R.}, \bibinfo{author}{Valtonen, E.},
  \bibinfo{author}{Santti, T.}, \bibinfo{author}{Pentik{\"{a}}inen, J.},
  \bibinfo{author}{Haeggstrom, E.}, \bibinfo{year}{2014}.
\newblock \bibinfo{title}{{Electronics for the RADMON instrument on the Aalto-1
  student satellite}}, in: \bibinfo{booktitle}{10th Eur. Work. Microelectron.
  Educ. EWME 2014}, pp. \bibinfo{pages}{161--166}.
\newblock \DOIprefix\doi{10.1109/EWME.2014.6877418}.
\bibitem[{Pierrard et~al.(2014)Pierrard, {Lopez Rosson}, Borremans, Lemaire,
  Maes, Bonnewijn, {Van Ransbeeck}, Neefs, Cyamukungu, Benck, Bonnet, Borisov,
  Cabrera, Gr{\'{e}}goire, Semaille, Creve, {De Saedeleer}, Desoete,
  Preud'homme, Anciaux, Helderweirt, Litefti, Brun, Pauwels, Quevrin, Moreau,
  Punkkinen, Valtonen, Hajdas and Nieminen}]{Pierrard-etal-2014}
\bibinfo{author}{Pierrard, V.}, \bibinfo{author}{{Lopez Rosson}, G.},
  \bibinfo{author}{Borremans, K.}, \bibinfo{author}{Lemaire, J.},
  \bibinfo{author}{Maes, J.}, \bibinfo{author}{Bonnewijn, S.},
  \bibinfo{author}{{Van Ransbeeck}, E.}, \bibinfo{author}{Neefs, E.},
  \bibinfo{author}{Cyamukungu, M.}, \bibinfo{author}{Benck, S.},
  \bibinfo{author}{Bonnet, L.}, \bibinfo{author}{Borisov, S.},
  \bibinfo{author}{Cabrera, J.}, \bibinfo{author}{Gr{\'{e}}goire, G.},
  \bibinfo{author}{Semaille, C.}, \bibinfo{author}{Creve, G.},
  \bibinfo{author}{{De Saedeleer}, J.}, \bibinfo{author}{Desoete, B.},
  \bibinfo{author}{Preud'homme, F.}, \bibinfo{author}{Anciaux, M.},
  \bibinfo{author}{Helderweirt, A.}, \bibinfo{author}{Litefti, K.},
  \bibinfo{author}{Brun, N.}, \bibinfo{author}{Pauwels, D.},
  \bibinfo{author}{Quevrin, C.}, \bibinfo{author}{Moreau, D.},
  \bibinfo{author}{Punkkinen, R.}, \bibinfo{author}{Valtonen, E.},
  \bibinfo{author}{Hajdas, W.}, \bibinfo{author}{Nieminen, P.},
  \bibinfo{year}{2014}.
\newblock \bibinfo{title}{{The Energetic Particle Telescope: First Results}}.
\newblock \bibinfo{journal}{\ssr} \bibinfo{volume}{184},
  \bibinfo{pages}{87--106}.
\newblock \DOIprefix\doi{10.1007/s11214-014-0097-8}.
\bibitem[{{Praks} et~al.(2018){Praks}, {Niemel{\"a}}, {N{\"a}sil{\"a}},
  {Kestil{\"a}}, {Jovanovic}, {Riwanto}, {Tikka}, {Leppinen}, {Vainio} and
  {Janhunen}}]{Praks-etal-2018}
\bibinfo{author}{{Praks}, J.}, \bibinfo{author}{{Niemel{\"a}}, P.},
  \bibinfo{author}{{N{\"a}sil{\"a}}, A.}, \bibinfo{author}{{Kestil{\"a}}, A.},
  \bibinfo{author}{{Jovanovic}, N.}, \bibinfo{author}{{Riwanto}, B.},
  \bibinfo{author}{{Tikka}, T.}, \bibinfo{author}{{Leppinen}, H.},
  \bibinfo{author}{{Vainio}, R.}, \bibinfo{author}{{Janhunen}, P.},
  \bibinfo{year}{2018}.
\newblock \bibinfo{title}{{Miniature Spectral Imager in-Orbit Demonstration
  Results from Aalto-1 Nanosatellite Mission}}, in:
  \bibinfo{booktitle}{Proceedings of 2018 IEEE International Geoscience and
  Remote Sensing Symposium}, pp. \bibinfo{pages}{1986--1989}.
\newblock \DOIprefix\doi{10.1109/IGARSS.2018.8517658}.
\bibitem[{{Shepherd}(2014)}]{Shepherd-2014}
\bibinfo{author}{{Shepherd}, S.G.}, \bibinfo{year}{2014}.
\newblock \bibinfo{title}{{Altitude-adjusted corrected geomagnetic coordinates:
  Definition and functional approximations}}.
\newblock \bibinfo{journal}{J. Geophys. Res. (Space Physics)}
  \bibinfo{volume}{119}, \bibinfo{pages}{7501--7521}.
\newblock \DOIprefix\doi{10.1002/2014JA020264}.
\bibitem[{{Th{\'e}bault} et~al.(2015){Th{\'e}bault}, {Finlay}, {Beggan},
  {Alken}, {Aubert}, {Barrois}, {Bertrand}, {Bondar}, {Boness} and
  {Brocco}}]{Thebault-et-al-2015}
\bibinfo{author}{{Th{\'e}bault}, E.}, \bibinfo{author}{{Finlay}, C.C.},
  \bibinfo{author}{{Beggan}, C.D.}, \bibinfo{author}{{Alken}, P.},
  \bibinfo{author}{{Aubert}, J.}, \bibinfo{author}{{Barrois}, O.},
  \bibinfo{author}{{Bertrand}, F.}, \bibinfo{author}{{Bondar}, T.},
  \bibinfo{author}{{Boness}, A.}, \bibinfo{author}{{Brocco}, L.},
  \bibinfo{year}{2015}.
\newblock \bibinfo{title}{{International Geomagnetic Reference Field: the 12th
  generation}}.
\newblock \bibinfo{journal}{Earth, Planets, and Space} \bibinfo{volume}{67},
  \bibinfo{pages}{79}.
\newblock \DOIprefix\doi{10.1186/s40623-015-0228-9}.
\bibitem[{{Vainio} et~al.(2009){Vainio}, {Desorgher}, {Heynderickx}, {Storini},
  {Fl{\"u}ckiger}, {Horne}, {Kovaltsov}, {Kudela}, {Laurenza} and
  {McKenna-Lawlor}}]{Vainio-etal-2009}
\bibinfo{author}{{Vainio}, R.}, \bibinfo{author}{{Desorgher}, L.},
  \bibinfo{author}{{Heynderickx}, D.}, \bibinfo{author}{{Storini}, M.},
  \bibinfo{author}{{Fl{\"u}ckiger}, E.}, \bibinfo{author}{{Horne}, R.B.},
  \bibinfo{author}{{Kovaltsov}, G.A.}, \bibinfo{author}{{Kudela}, K.},
  \bibinfo{author}{{Laurenza}, M.}, \bibinfo{author}{{McKenna-Lawlor}, S.},
  \bibinfo{year}{2009}.
\newblock \bibinfo{title}{{Dynamics of the Earth's Particle Radiation
  Environment}}.
\newblock \bibinfo{journal}{\ssr} \bibinfo{volume}{147},
  \bibinfo{pages}{187--231}.
\newblock \DOIprefix\doi{10.1007/s11214-009-9496-7}.

\end{thebibliography}

\end{document}